\documentclass[
reprint,
superscriptaddress,
showkeys,
showpacs,
amsmath,amssymb,
aps,
pra,
]{revtex4-1}

\usepackage{graphicx}
\usepackage{sidecap}
\usepackage{dcolumn}
\usepackage{upgreek}
\usepackage{mathrsfs}
\usepackage{setspace}
\usepackage{bm}
\usepackage[breaklinks=true,colorlinks=true,linkcolor=blue,urlcolor=blue,citecolor=blue]{hyperref}
\allowdisplaybreaks[4]
\begin{document}
\preprint{APS/123-QED}

\title{A concise review of Rydberg atom based quantum computation and quantum simulation}

\author{Xiaoling Wu}
\affiliation{State Key Laboratory of Low Dimensional Quantum Physics, Department of Physics, Tsinghua University, Beijing 100084, China}

\author{Xinhui Liang}
\affiliation{State Key Laboratory of Low Dimensional Quantum Physics, Department of Physics, Tsinghua University, Beijing 100084, China}

\author{Yaoqi Tian}
\affiliation{School of Physical Sciences, University of Chinese Academy of Sciences, Beijing 100049, China}
\affiliation{CAS Key Laboratory of Theoretical Physics, Institute of Theoretical Physics, Chinese Academy of Sciences, Beijing 100190, China}

\author{Fan Yang}
\email{fanyangphys@gmail.com}
\affiliation{State Key Laboratory of Low Dimensional Quantum Physics, Department of Physics, Tsinghua University, Beijing 100084, China}

\author{\\Cheng Chen}
\affiliation{State Key Laboratory of Low Dimensional Quantum Physics, Department of Physics, Tsinghua University, Beijing 100084, China}

\author{Yong-Chun Liu}
\affiliation{State Key Laboratory of Low Dimensional Quantum Physics, Department of Physics, Tsinghua University, Beijing 100084, China}
\affiliation{Frontier Science Center for Quantum Information, Beijing 100084, China}

\author{Meng Khoon Tey}
\affiliation{State Key Laboratory of Low Dimensional Quantum Physics, Department of Physics, Tsinghua University, Beijing 100084, China}
\affiliation{Frontier Science Center for Quantum Information, Beijing 100084, China}

\author{Li You}
\email{lyou@mail.tsinghua.edu.cn}
\affiliation{State Key Laboratory of Low Dimensional Quantum Physics, Department of Physics, Tsinghua University, Beijing 100084, China}
\affiliation{Frontier Science Center for Quantum Information, Beijing 100084, China}
\affiliation{Beijing Academy of Quantum Information Sciences, Beijing 100193, China}

\begin{abstract}
	 Quantum information processing based on Rydberg atoms emerged as a promising direction two decades ago. Recent experimental and theoretical progresses have shined exciting light on this avenue. In this concise review, we will briefly introduce the basics of Rydberg atoms and their recent applications in associated areas of neutral atom quantum computation and simulation. We shall also include related discussions on quantum optics with Rydberg atomic ensembles, which are increasingly used to explore quantum computation and quantum simulation with photons.
\end{abstract}



\maketitle

\section{Introduction}\label{sec:sec1}
Rydberg atoms possess highly excited valance electrons far away from atomic cations \cite{stebbings1983rydberg,gallagher2005rydberg}. Compared with ground state atoms, they exhibit exaggerated properties such as enormously large electric dipole moments, which can facilitate strong interactions with macroscopic external fields or even microscopic electromagnetic fields from nearby particles. These interactions can be controlled by static electric or magnetic, laser, or microwave fields, making systems of Rydberg atoms ideal choices for implementing controllable quantum many-body simulators. Aided by impressive experimental progresses in neutral atom based systems during the past few decades, in areas including preparations of ultracold atomic gases \cite{phillips1998nobel,metcalf2007laser}, high-resolution imaging of single atoms \cite{sherson2010single,bakr2010probing}, and trapping of individual atoms in reconfigurable optical tweezer arrays \cite{kim2016situ,endres2016atom,barredo2016atom}, the appealing features  of highly excited Rydberg states are convincingly revealed which establish them as the most popular neutral atom based platforms for quantum information processing (QIP).

A large body of QIP concerns quantum computation and quantum simulation, which aims at solving complex problems otherwise intractable or difficult for classical computers. The breadth of physical candidates being pursued for implementing quantum computation and quantum simulation includes neutral atoms, superconducting circuits \cite{wendin2017quantum,huang2020superconducting}, semiconductor heterostructures \cite{eriksson2004spin,zhou2016semiconductor,jazaeri2019review}, trapped ions \cite{haffner2008quantum,bruzewicz2019trapped}, or linear optics \cite{kok2007linear,slussarenko2019photonic,sharma2019review},  etc. Neutral atom based qubits are of particular interest because of their nature born properties of being uniformly identical, easy to scale up, long coherence time, and their potentials for versatile direct interactions mediated by Rydberg states, which are difficult to match in other systems or implementations. In addition, strong Rydberg interactions can be mapped onto interactions between individual photons, opening up the possibilities of performing photonic quantum computation and quantum simulation in a deterministic manner, which are on top of the distinct advantages of transmitting photonic quantum information to far away locations and of exchanging quantum information coded in photons to atoms or material qubits.

In this concise review, we attempt to present an up-to-date status report of Rydberg atom based quantum computation and quantum simulation, focusing on the basic mechanisms as well as recent progresses, but at the same time striving to be as complete as we can in this fast-developing field. Earlier results on selected topics can be found in Refs.~\cite{saffman2010quantum,saffman2016quantum,firstenberg2016nonlinear,murray2016quantum}, while more details on recent progresses are documented and reviewed in Refs.~\cite{browaeys2020many,henriet2020quantum,morgado2020quantum}. Our review will be divided into the following sections. In Sec.~\ref{sec:sec2}, we introduce the basic properties of Rydberg atoms. The current state of the art for quantum computation and quantum simulation with Rydberg atoms follows in Sec.~\ref{sec:sec3} and Sec.~\ref{sec:sec4}, respectively. In Sec.~\ref{sec:sec5}, we discuss Rydberg atomic ensemble based quantum optics studies, which further support explorations of QIP with photons. We conclude in Sec.~\ref{sec:sec6} with a summary and an outlook into the future.

\section{Basic properties of Rydberg atoms}
\label{sec:sec2}
Research on Rydberg atoms can be traced back to more than 100 years ago. As early as 1885, when Johann Balmer was studying the spectrum of hydrogen atoms, he discovered a simple formula that could describe a series of spectral lines \cite{white1934introduction}. In 1888, the Swedish physicist Johannes Rydberg extended the Balmer formula to a more general Rydberg formula \cite{foot2005atomic}
\begin{equation}
\frac{1}{\lambda}=R_{\rm H}\left( \frac{1}{n^{2}}-\frac{1}{m^{2}} \right), \label{eq:eq1}
\end{equation}
where $\lambda$ is wavelength, $R_{\rm H}=1.097\times 10^{7} \rm m^{-1}$ is the Rydberg constant of hydrogen atom, and $n$, $m$ are integers, known as principal quantum numbers later. This formula played an important role in understanding and establishing quantum mechanics \cite{semat2012introduction}. More generally, any atom (molecule or semiconductor quantum dot) in a state with a highly-excited electron, i.e., with valance electron in a large principal quantum number $n$ state of several tens to hundreds or even higher, is regarded as Rydberg atom.
\subsection{Universal properties}\label{subsec:sub2.1}
The electron in a Rydberg atom is in a large weakly bound orbit, which makes Rydberg atoms exhibit incredibly exaggerated properties relative to ground state ones \cite{gallagher2005rydberg,gounand1979calculation}. In general, almost all properties of Rydberg atoms show a characteristic scaling with $n$, or more precisely the effective principal quantum number $n^{*}=n-\delta_{nlj}$, where $\delta_{nlj}$ denotes quantum defect \cite{seaton1958quantum} which generally could depend on principal quantum number $n$, orbital angular momentum quantum number $l$, and total angular momentum quantum number $j$. Typical scaling laws for a list of common properties are summarized in Table~\ref{tab:table1}. 

\begin{figure}
	\centering
	\includegraphics[width=\linewidth]{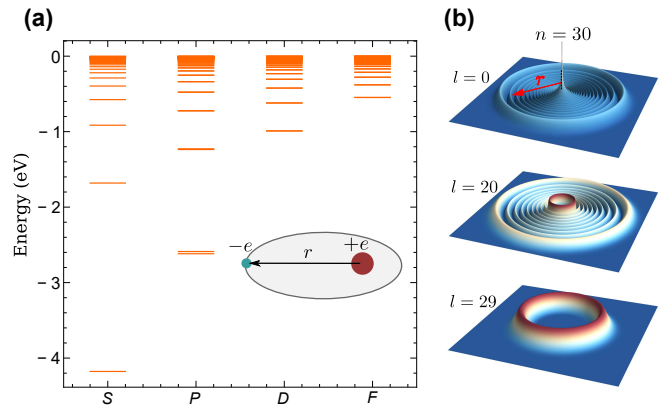}
	\caption{Energy levels and electron probability density of $^{87}$Rb atom. Relevant data and calculations are based on Alkali Rydberg Calculator (ARC) Toolbox \cite{vsibalic2017arc}. (a) Distribution of energy levels for $5\leq n \leq 60$, $0\leq l \leq 3$. (b) Density distributions of Rydberg electron in the radial direction ($r|\psi(r)|^2$). The bottom figure shows that the circular Rydberg atom ($l=n-1$) is highly localized around $n^2{a_0}$.}
	\label{fig:fig2_2_1}
\end{figure}

For instance, Fig.~\ref{fig:fig2_2_1}(a) plots the energy levels of $^{87}$Rb atom. The distribution of energy levels becomes denser as $n$ increases because the energy difference $\Delta E$ between neighboring Rydberg states scales as $n^{-3}$, on the order of GHz when $n$ is around 100. The size of Rydberg atoms scales as $\langle r \rangle \propto n^2 a_{0}$ with $a_{0}$ the Bohr radius, indicating that the dipole moment of Rydberg atoms also scales as $n^2$. Select plots for probability distributions of Rydberg-electron radial wavefunction $\psi(r)$ in the equatorial plane for states with different $l$ are shown in Fig.~\ref{fig:fig2_2_1}(b). It oscillates with $r$, and the oscillation frequency decreases as $l$ increases. Especially noteworthy is, when $l$ takes the maximal value $l=n-1$, the electron is localized around $r=n^2{a_0}$, forming a nearly circular orbit and such atoms are called circular Rydberg atoms.

\begin{table}[b]
	\caption{\label{tab:table1}%
		Scaling with $n$, the principle quantum number, for the listed properties of Rydberg atoms.
	}
	\begin{ruledtabular}
		\begin{tabular}{lcc}
			\multicolumn{1}{l}{\textrm{properties}}& 
			\multicolumn{1}{c}{\textrm{notation}}&
			\multicolumn{1}{c}{\textrm{$\beta$(scaling $n^\beta$)}}\\ 
			\hline
			orbit size                                   & $\langle r\rangle $                 & 2                \\
			binding energy                                 & $E_{n}$             & -2               \\
			neighboring level spacing                               & $E_{n+1}-E_{n}$ & -3               \\
			radiative lifetime                             & $\tau_{0}$            & 3                \\
			dipole matrix element\footnote{$g$ denotes ground state}                          & $\langle g|er|nl\rangle $          & -3/2             \\ 
			dipole matrix element\footnote{dipole matrix elements between neighbouring Rydberg states} & $\langle  nl|er|nl+1\rangle $    & 2                \\
			polarizability                                 & $\alpha$              & 7                \\
		\end{tabular}
	\end{ruledtabular}
\end{table}

\subsection{Lifetime}\label{subsec:sub2.2}
Essentially, all Rydberg atoms are unstable. They decay with a characteristic lifetime $\tau$ according to the states they reside in, and eventually can fall to the ground state. Two physical mechanisms dominate the decay process, one is spontaneous emission caused by perturbation of the vacuum electromagnetic fluctuations at $0$~K (with a rate $\tau_0$), the other is stimulated emission caused by blackbody radiation (BBR) at finite temperature $T$ (with a rate $\tau_{bb}$). The total lifetime $\tau$ of a Rydberg atom at finite temperature is given by
\begin{equation}
\frac{1}{\tau}=\frac{1}{\tau_{0}}+\frac{1}{\tau_{bb}},\label{eq:eq2}
\end{equation}

{\it Radiative lifetime}.---$\tau_{0}$ of a given Rydberg state $n$ can be calculated by summing up the transition rates of all possible spontaneous emission channels, i.e., 
\begin{equation}
\frac{1}{\tau_{0}}=\sum_{n^\prime}^{}A_{nn^\prime},\label{eq:eq3}
\end{equation}
where $A_{nn^\prime}$ is the Einstein $A$ coefficient for the transition from state $n$ to a lower one $n'$ \cite{gallagher1979interactions}. From the perspective of classical electrodynamics, the farther an electron is away from the nucleus, the smaller acceleration it experiences and therefore the smaller radiation rate is. Roughly speaking the lifetime of Rydberg atom increases as the principal quantum number $n$ becomes larger. More rigorous analysis shows that the scaling law between $\tau_{0}$ and $n$ for different $l$ states are not the same \cite{beterov2009quasiclassical}. For state $|nl\rangle$ with a low orbital quantum number $l$, there exists a large number of dipole allowed transition channels. The decay rates scale as $n^{-5}$ or $n^{-3}$ \cite{beterov2009quasiclassical} for transitions to neighboring Rydberg states or to low-energy states, respectively. The transitions to low-lying states constitute dominant channels of spontaneous emission, so the radiative lifetime is approximately proportional to $n^3$. However, the situation will be different if the atom is in a high-$l$ state, e.g., circular Rydberg state with $l=n-1$, whose dominant dipole allowed final state is $|n^\prime=n-1,l^\prime=n-2\rangle$. The radiative lifetime of such states is proportional to $n^{5}$, which can be much larger than that of low-$l$ states, making circular Rydberg atoms prospective  carriers for quantum information \cite{hulet1983rydberg,facon2016sensitive,signoles2017coherent,nguyen2018towards,aliyu2018transport,cortinas2020laser}.

{\it Blackbody lifetime}.---Under normal circumstances, the effects of room-temperature blackbody radiation on atomic systems can be
neglected, because for atoms in the ground and low excited states with large transition frequencies $\omega$, the effective
number $n_{\omega}$ of BBR photons at $T=300$~K is much smaller than unity \cite{beterov2009quasiclassical}. This effect, however, becomes more dramatic for atoms in Rydberg states because these states have low-frequency transitions with large electric dipole matrix elements and are thus strongly coupled to the BBR \cite{gallagher1979interactions}. Different from spontaneous emission transitions, BBRs mainly cause atoms to decay to nearby Rydberg states, because the lower the transition frequencies, the higher the mean number of photons in near resonant modes. When the principal quantum number $n$ is large, we only need to focus on the transitions with $\omega\ll k_{B}T$, whose lifetimes are approximately given by $\tau_{bb}=3\hbar n^{2}/4\alpha^{3}k_{B}T$ ($\hbar=1$ is assumed in the rest of this paper), with $\alpha$ the fine structure constant. This scaling of $\tau_{bb}\propto n^{2}$ on $n$ is the same for different $l$ states. Taking $^{87}$Rb atom at room temperature ($T=300$~K) as an example, for Rydberg states with $n\sim 50$, the total lifetime is on the order of $50$~$\mu$s, which suggests MHz-rate  operations are required to ensure quantum coherence.

{\it Superradiance}.---The decay mechanisms mentioned above apply to every single excited atom, which is usually a valid scenario for experiments in dilute gases. The total lifetime, however, would decrease in dense gases due to collective and cooperative effects caused by photon exchange interaction between different Rydberg atoms. The enhanced decay rates mostly arise from superradiance \cite{dicke1954coherence}, where photons emitted by excited atoms can cause stimulated emission in other atoms, triggering an avalanche of collective decays. Such superradiant signature has been observed experimentally \cite{wang2007superradiance}.

\subsection{Rydberg-Rydberg interaction}\label{subsec:sub2.3}
Since Rydberg atoms possess large dipole moments scaling as $n^{2}$, the resulting dipole-dipole interactions increase quickly as $n$ increases. One of the main advantages of Rydberg interaction lies at its versatile controllability, the flexibilities of controlling its strength, sign, anisotropy, and spatial dependence by choosing appropriate states, supplemented by the ability to turn off the interaction by transferring atoms back to ground state \cite{pritchard2012cooperative}. In the following, we will provide a brief introduction of Rydberg-Rydberg interaction and its associated extensions.

\subsubsection{Dipole-dipole Interactions}
Rydberg atom can be regarded as an electric dipole composed of its highly excited electron and its cation core, whose dipole moment is $\bm p =-e{\bm d}$, where $\bm d$ is the relative displacement from the cation to the Rydberg electron. The corresponding dipole-dipole interaction between two Rydberg atoms can be expressed as
\begin{equation}
V_\mathrm{dd}=\frac{e^{2}}{4\pi \epsilon_{0}} \frac{ {\bm{d}_{1}} \cdot {\bm{d}}_{2}- 3 ({\bm{d}_{1}} \cdot  {\bm{e}}_R)({\bm{d}_{2}} \cdot {\bm{e}}_R)}{R^{3}}, \label{eq:eq4}
\end{equation}
where ${\bf e}_R$ is the unit vector along the relative coordinate ${\bm R}$ between two atoms, and $R=|{\bm R}|$ is the interatomic distance [see Fig.~\ref{fig:fig2_2}(a)]. Such dipolar interaction scales as $V_\mathrm{dd}\propto n^4$ because the dipole moment scales as $|{\bm d}|\propto n^2$. When there exists no external electric field, atoms are unpolarized, the spatial symmetry of wavefunction results in a vanishing static dipole moment, i.e., $\langle{\bm d}\rangle = 0$, and thus $\langle V_\mathrm{dd}\rangle =0$. However, the dipole operator $\bm d$ has non-vanishing  matrix elements between eigenstates with different parities. The dipole-dipole interaction matrix $\hat{V}_\mathrm{dd}$ should therefore be introduced to accurately describe the interaction between different Rydberg states. 

In spherical hamonic basis, dipole-dipole interaction operator can be expressed as
\begin{equation}
\hat{V}_\mathrm{dd}=\frac{-e^2}{4\pi \epsilon_{0} R^3}\sqrt{\frac{24\pi}{5}}\sum_{\mu,\nu}^{}C_{\mu,\nu,\mu+\nu}^{1,1,2}Y_{2}^{\mu+\nu}(\theta,\phi)^{*}\hat{d}_{\mu}^{(1)}\hat{d}_{\nu}^{(2)},\label{eq:eq5}
\end{equation}
where $C_{m_1,m_2,M}^{j_1,j_2,J}$ is Clebsch-Gordan coefficient, $Y_{l}^{m}(\theta,\phi)$ is spherical harmonic function, and $\hat{d}_{q}^{(i)}$ denotes the spherical component $(q=\pm1,0)$ of the dipole operator for the $i$th atom. For simplicity, we consider pairwise case with atoms labeled as $1$ and $2$ and a single transition channel $r_{1}+r_{2}\rightarrow r^\prime_{1}+r^\prime_{2}$ [see Fig.~\ref{fig:fig2_2}(b)]. The Hamiltonian of the model system in the subspace spanned by bases $\{  {|r_{1}r_{2}\rangle} ,{|r^\prime_{1}r^\prime_{2}\rangle} \}$ can be written as
\begin{equation}
\hat{H}=\begin{bmatrix} 0 & {C_{3}(\theta)}/{R^{3}} \\ {C_{3}(\theta)}/{R^{3}} & \delta_{F}
\end{bmatrix}, \label{eq:eq6}
\end{equation}
with $C_{3}(\theta)\propto n^4$ the anisotropic interaction coefficient, and $\delta_{F}=(E_{r^\prime_{1}}+E_{r^\prime_{2}})-(E_{r_{1}}+E_{r_{2}})$ the F{\"o}ster defect defined as the difference of the bare energy between final and initial pair states. In Eq.~(\ref{eq:eq6}), we neglect the dependence of $\hat{H}$ on the azimuth angle $\phi$, since it only contributes a phase factor to the interaction strength $V = C_{3}(\theta)/R^{3}$ for transition between well-defined quantum numbers. Other channels such as $r_{1}+r_{2}\rightarrow r^{\prime\prime}_{1}+r^{\prime\prime}_{2}$ shown in Fig.~\ref{fig:fig2_2}(b) can be neglected due to their relatively larger F\"{o}ster defects.

The eigenvalues of $\hat{H}$ are given by
\begin{equation}
E_{\pm}=\frac{\delta_{F}}{2}\pm \frac{1}{2}\sqrt{\delta_{F}^2+4V^2},\label{eq:eq7}
\end{equation}
from which we can analyze the $R$-dependence of the interaction induced energy shift $\Delta E_{\pm} = E_\pm(R)-E_\pm(R\rightarrow\infty)$. For $\delta_{F}=0$, the bare energies of the initial and the final state are degenerate, in this case the off-diagonal element $V$ can lead to strong coupling between the initial and final pair (two-atom) states. As a result, the eigenstates become $|\pm\rangle=\left(|{r_{1}r_{2}}\rangle\pm|{r'_{1}r'_{2}}\rangle\right)/\sqrt{2}$, whose eigenenergies are proportional to $\pm C_{3}/R^{3}$. This so-called F\"{o}ster resonance has been observed in several Rydberg systems \cite{safinya1981resonant,pillet2009controllable,ravets2014coherent}. In particular, if $r_1=r'_2=r,r'_1=r_2=r'$, the initial state $|rr'\rangle$ and final state $|r'r\rangle$ are symmetric with respect to exchange of the two atoms, which naturally satisfies F\"{o}ster resonance condition $\delta_F=0$. The concept of F\"{o}ster resonance can also be generalized to three-body and many-body situations \cite{faoro2015borromean,gurian2012observation,ryabtsev2018coherence}. For $\delta_{F}\neq 0$, the interaction induced level shift exhibits a crossover with the increase of interatomic distance $R$: when two atoms are so close that $\delta_{F}\ll V(R)$, the off-diagonal elements of $\hat{H}$ are dominant, similar to the case of F\"{o}ster resonance, and $\Delta E_{\pm} \approx \pm C_{3}/R^{3}$; while if $\delta_{F}\gg V(R)$, the interaction terms can be treated as a perturbation that just slightly shifts the energies of bare states, giving rise to $\Delta E_-=-C_{6}/R^6$ for $|r_{1}r_{2}\rangle$ and $\Delta E_{+}=C_{6}/R^6$ for $|r^\prime_{1}r^\prime_{2}\rangle$ with $C_{6}={C_{3}}^2/\delta_{F}$. Interaction in this regime is actually proportional to $R^{-6}$, which is conventionally called van der Waals (vdW) interaction. To describe the crossover of the energy shift from $R^{-3}$ to $R^{-6}$, a characteristic radius $R_\mathrm{vdW}={|C_{6}/\delta_{F}|}^{{1}/{6}}$ is usually defined, at which $V(R_\mathrm{vdW})=\delta_{F}$.

In a more elaborate treatment, all Rydberg states close to the initial state have to be taken into consideration, which usually result in complicated interaction potential curves. Without F\"{o}ster resonance, interaction at large distance $R$ can be treated perturbatively as mentioned above, but the resulting vdW interaction matrix contains both diagonal and off-diagonal elements. The calculation of Rydberg interaction becomes more complicated, and the details can be found in Ref.~\cite{weber2017calculation}.

\subsubsection{Rydberg blockade}
Due to the strong mutual interaction between Rydberg atoms, within a certain volume in an ensemble, only one atom can be excited from the ground state to Rydberg state, as the first excited Rydberg atom shifts the Rydberg energy levels of all other nearby atoms off resonance. This so-called Rydberg blockade phenomenon facilitates conditional dynamics highly desired for QIP.
\begin{figure}
	\centering
	\includegraphics[width=\linewidth]{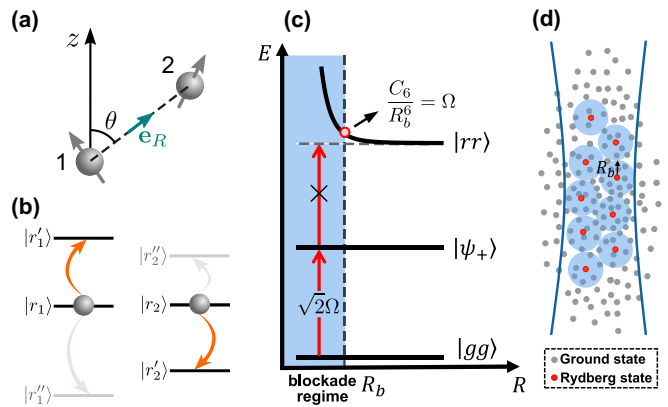}
	\caption{Rydberg interaction and Rydberg blockade. (a) Dipole-dipole interaction between two atoms with interatomic separation $R$ at an angle $\theta$ to the quantization axis $z$. (b) Relevant energy levels and corresponding transition channels contributing to the dipole-dipole interaction. (c) Rydberg blockade in the two-atom case. (d) Rydberg superatoms in an ensemble of atoms. All atoms inside a volume with radius $R_{b}$ share a single Rydberg excitation.}
	\label{fig:fig2_2}
\end{figure}

To illustate the principle of Rydberg blockade, we first consider the simplest case of two-atoms, as shown in Fig.~\ref{fig:fig2_2}(c). The resonant excitation light can drive two atoms from ground state $|gg\rangle$ to symmetric single-atom excited state $|\psi_+\rangle=(|gr\rangle+|rg\rangle)/\sqrt{2}$. The doubly excited state $|rr\rangle$, however, is shifted to off-resonant energy particularly if the distance between atoms satisfies $R<R_{b}$, where $R_{b}$ is a critical distance called Rydberg blockade radius. As a result, one cannot excite two atoms simultaneously to Rydberg states if they are located sufficiently close-by to each other. Rydberg blockade in the two-atom system is independently demonstrated by Urban \emph{et al.} \cite{urban2009observation} and Ga{\"e}tan \emph{et al.} \cite{gaetan2009observation}, and more recently by Zeng \emph{et al.} \cite{Zeng2017}.

If we consider a volume with radius $R_{b}$ containing $N$ atoms, the singly excited state refers to the situation when all atoms in this volume share a single atomic excitation, which is described as a coherent superposition state
\begin{equation}
|W\rangle=\frac{1}{\sqrt{N}}\sum_{i=1}^{N}|g_{1}g_{2}\cdots r_{i}\cdots g_{N}\rangle, \label{eq:eq8}
\end{equation}
where $i$ labels the excited atom. This state is often referred to as a ``superatom'' [see Fig.~\ref{fig:fig2_2}(d)]. The Rabi frequency between the ground state $|G\rangle=|g_{1}g_{2}\cdots g_{N}\rangle$ and such a collective excited state $|W\rangle$ is enhanced to $\Omega=\sqrt{N}\Omega_{1}$, with $\Omega_{1}$ the corresponding single-atom Rabi frequency \cite{lukin2001dipole}. The evidence of this collective effect is first observed by Heidemann \emph{et al.} \cite{heidemann2007evidence} and further confirmed by Dudin \emph{et al.} based on coherent many-body Rabi oscillations \cite{dudin2012observation}. More details about Rydberg blockade can be found in Ref.~\cite{comparat2010dipole}.

Interaction between Rydberg atoms can give rise to other collective phenomena, such as antiblockade. In contrast to Rydberg blockade, two or more atoms can conditionally be simultaneously excited to Rydberg states in the antiblockade regime, where the interaction induced level shift is compensated by the detuning of the driving light \cite{ates2007antiblockade,amthor2010evidence}.

\subsubsection{Rydberg dressing}
In addition to Rydberg blockade, off-resonant laser or microwave dressing of Rydberg states provide alternative routes towards realizing tunable interactions.

\begin{figure}
	\centering
	\includegraphics[width=\linewidth]{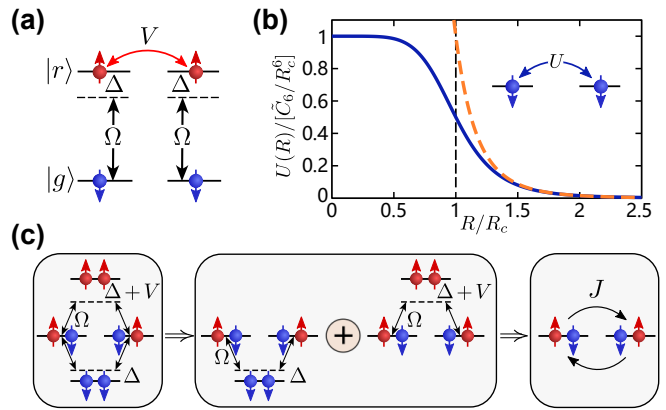}
	\caption{Ground-state Rydberg dressing. (a) Two atoms are dressed by a laser of Rabi frequency $\Omega$ and detuning $\Delta$, and interact with each other via a strength $V$ when they are both in the Rydberg state $|r\rangle$. (b) Effective induced interaction $U(R)$ between two dressed ground-state atoms (blue solid line) and its vdW type tail (orange dashed line). (c) Mechanism for inducing a synthetic spin-exchange interaction between the ground state and the Rydberg state, where the vdW interaction $V$ breaks the symmetry of the two opposing perturbation paths.}
	\label{fig:fig2_3}
\end{figure}

Off-resonant laser dressing can admix a small fraction of Rydberg state $|r\rangle$ to the ground state $|g\rangle$, giving rise to a dressed eigenstate
$|\psi\rangle\sim |g\rangle+\varepsilon|r\rangle$. The Rydberg state fraction is controlled by the dressing parameter $\varepsilon\propto(\Omega/2\Delta)$, where $\Omega$ and $\Delta$ are Rabi frequency and detuning of the dressing field, respectively [see Fig.~\ref{fig:fig2_3}(a)]. The decay rate of this dressed state is reduced to $\sim \varepsilon^{2}\gamma_{r}$ with $\gamma_{r}$ the decay rate for the Rydberg state. Under realistic experimental conditions, such a dressed state permits coherent evolution of the system over much longer time. Assuming atoms in states $|r\rangle$ interact with each other via a vdW interaction $C_6/R^6$, for $C_{6}\Delta>0$, two dressed ground-state atoms acquire a modified soft-core potential given by \cite{henkel2010three,johnson2010interactions} 
\begin{equation}
U(R)=\frac{\tilde{C_{6}}}{R^6+R_{c}^6},\label{eq:eq9}
\end{equation}
with $\tilde{C_{6}}=(\Omega/2\Delta)^4 C_{6}$ the effective interaction coefficient and $R_c=(C_{6}/2\Delta)^{1/6}$ a critical radius inside which the actual interaction begins to deviate from the simple vdW type significantly. As shown in Fig.~\ref{fig:fig2_3}(b), the peculiar shape of $U(R)$ can be understood as follows: when two atoms are separated by a distance $R\gg R_{c}$, dressing of two atoms is not influenced by atom-atom interaction, leading to an admixed vdW type interaction $\sim \varepsilon^4 C_6/R^6$ between them. At smaller distance $R\ll R_{c}$, however, the effective potential approaches a constant, because the double excitation component $\varepsilon^2|rr\rangle$ is strongly suppressed due to Rydberg blockade. When $C_{6}$ and $\Delta$ are opposite in sign, $U(R)$ is altered to a resonance-shape interaction potential due to antiblockade \cite{balewski2014rydberg}.

The above discussed Rydberg dressing induced interaction is measured by Jau \emph{et al}. in a two-atom system \cite{jau2016entangling} and by Zeiher \emph{et al}. in a many-body system \cite{zeiher2016many}. Subsequently, this dressing scheme is employed to study coherent many-body dynamics of atoms in an optical lattice \cite{zeiher2017coherent} and quantum phase transition in an atomic ensemble \cite{borish2020transverse}. It is proposed that laser dressing can give rise to a richer variety of interactions when more ground states are coupled to different Rydberg states \cite{wuster2011excitation,glaetzle2015designing,van2015quantum}. Furthermore, the dressing scheme can be extended to establish an effective interaction between a ground-state atom and a Rydberg atom \cite{glaetzle2017quantum,letscher2018mobile,yang2019quantum}. For example, Yang, Yang, and You propose a scheme to engineer spin-exchange interaction that can coherently transfer Rydberg excitation in one atom to nearby ground-state atoms \cite{yang2019quantum} [see Fig.~\ref{fig:fig2_3}(c)].

Optical dressing scheme can be realized in a three-level configuration as well, which can potentially outperform the two-level scheme in terms of interaction strength and lifetime \cite{helmrich2016two,gaul2016resonant}. Different from the laser dressing scheme that couples a ground state to a Rydberg state, microwave dressing induces mixing of different Rydberg states close to each other, which facilitates tuning of their original interaction forms into more desirable ones. For example, interaction between atoms in microwave-dressed Rydberg states can be adjusted from vdW type to F\"{o}ster resonance \cite{bohlouli2007enhancement,sevinccli2014microwave,tretyakov2014controlling}. Under suitable parameters, interaction can even be turned off completely \cite{shi2017annulled,young2020asymmetric}. Microwave dressing can also reduce the sensitivity of the dynamics to interatomic mechanical forces \cite{petrosyan2014binding} or to stray static electric fields \cite{booth2018reducing}.

\section{Quantum Computation with Rydberg atoms}\label{sec:sec3}
\begin{figure*}
	\centering
	\includegraphics[width=\linewidth]{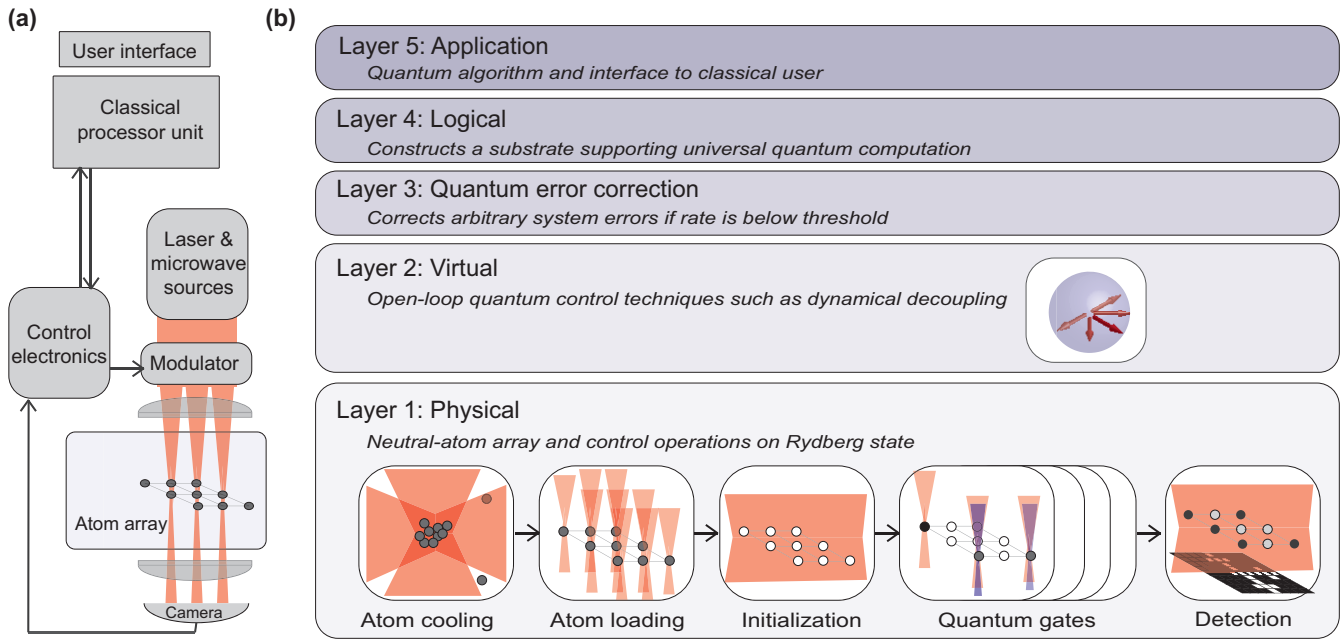}
	\caption{Neutral atom quantum computation platform. (a) The platform consists of a classical computer and a neutral atom based quantum processor. The later is composed of atom arrays in a vacuum chamber, peripherals for detection and control of atoms, e.g., laser/microwave resources, optical/microwave modulators, cameras and corresponding control electronics. (b) A generic outline for neutral atom quantum computation architecture (see Ref.~\cite{PhysRevX.2.031007}). In the physical layer, neutral atoms in the quantum processor are first cooled and captured by a magnetic optical trap (MOT) in ultra-high vacuum chamber. After subsequently loading atoms into an optical tweezer array or an optical lattice, they are initialized with optical pumping, followed by a sequence of quantum gates controlled via laser and microwave fields. In the end, quantum information is read out through fluorescence imaging. The combination of noisy physical qubits and open-loop quantum control techniques such as dynamical decoupling constructs virtual qubits with improved effective coherence time and minimized systematic gate errors, which serve as the basic building blocks for quantum-error-correction-based universal quantum computation or noisy intermediate-scale quantum applications. In the logic and QEC layer, decompositions of quantum algorithm in terms of quantum circuits, together with data processing of detection results, are processed by classical computer. The user interfaces in the application layer are also aided by classical computer.}
	\label{fig:fig3_1}
\end{figure*}

In this section, we follow the layered quantum-computer architecture which decomposes a complex quantum information processing system into a manageable set of subsystems~\cite{PhysRevX.2.031007} to provide a sense of how rudimentary neutral atom quantum computation platform (shown in Fig.~\ref{fig:fig3_1}) works.

In the physical layer, which we mostly focus on in this review, atoms evolve with sequences of controls, e.g., atom cooling and loading, qubit initialization and detection, quantum gates, loss detection and reloading. 
In the virtual layer, open-loop quantum control techniques such as dynamical decoupling are used to increase coherence time. Essential requirements in the first two layers are embodied by the DiVincenzo criteria~\cite{divincenzo2000physical}. Basic principles and recent progresses are briefly summarized in subsections ~\ref{subsec:sub3.1}-\ref{subsec:sub3.4}.

Limited by demonstrated fidelities of quantum gates operations, few theoretical studies are reported in the quantum error correction (QEC) layer or beyond~\cite{crow2016improved, auger2017blueprint}. Nevertheless, some near-term applications can still be impelmented in the regime of noisy intermediate-scale quantum (NISQ) operations, which will be briefly discussed in subsection \ref{subsec:sub3.5}.

\subsection{Physical qubit and its coherence time}
\label{subsec:sub3.1}
In theory, any quantum system, with more than two distinguishable states, can be used to encode a quantum-bit of information. Neutral atoms, similar to trapped ions, from which come a variety of species and quantum states to choose, provide abundant yet identical choices of physical qubits with rich internal quantum states. The different choices of atoms and states are mainly decided based on trade-offs between well isolated states possessing longer coherence times and easily accessible quantum levels for convenient initializations, manipulations, detections, and implementations of quantum gates.

Aided by the developed techniques for laser cooling and optical or magnetic trapping, heavy alkali atoms like Rb and Cs are investigated in earlier experiments~\cite{saffman2010quantum}. Hyperfine ground states are used to encode qubits due to their long coherence times and large level spacings up to GHz regime. Moreover, they can be readily initialized and detected with optical pumping and resonance fluorescence. Recently, more atomic species are investigated and new possibilities for higher fidelity qubit operations are demonstrated. Alkali earth atoms like Sr~\cite{madjarov2020high}, with two valence electrons, can be further cooled in an optical trap by narrow-line cooling, leading to further suppressed decoherence caused by atomic motion~\cite{cooper2018alkaline}. One photon excitation from ground clock states to Rydberg levels usually offers a higher excitation Rabi frequency than two-photon Rydberg excitations of alkali atoms at similar level of laser powers~\cite{madjarov2020high}. 

Long coherence time of physical qubits is one of the basic requirements for quantum computation. More than ${10}^{4}$ quantum gates within coherence time are needed to meet the threshold of QEC~\cite{crow2016improved}. Detailed analysis of decay mechanisms of the hyperfine coherence of trapped atoms is carried out by Kuhr {\it et al}.~\cite{kuhr2005analysis}. Under nominal circumstances, atom loss rate is dominated by collisions with background gases and worsened by heating of atoms~\cite{gehm1998dynamics}. With proper shielding of radio-frequency (RF) noise, longitudinal relaxation time is constrained by Raman scattering, which can be suppressed by confining atoms at intensity minimum of blue detuned optical trap~\cite{li2012crossed}. Transverse relaxation time is affected by a host of dephasing mechanisms: including fluctuation of magnetic field, fluctuation of trap depth and position, while thermal motion of atoms can be suppressed by cooling atoms to motional ground states~\cite{thompson2013coherence, wu2019stern}. Coherence time of hundreds of milliseconds without dynamical decoupling is achieved aided by the technique of magic-intensity trapping~\cite{yang2016coherence}, which nulls out the first order dependence of energy level on trapping light-intensity. A similar technique is used to balance coherence times of mixed isotopes and further prolongs coherence time close to 1 s~\cite{guo2020balanced}.

When building up a two-qubit quantum gate, Rydberg excitation or dressing is employed, which also demands a long coherence time. Imperfections in coherent optical excitation of Rydberg states are analyzed in detail in Ref.~\cite{de2018analysis}. Major contributions to decoherence come from finite lifetime due to spontaneous emission and BBR caused transitions. Coherence time up to 27~$\mu$s is demonstrated in experiment helped by reduced laser phase noise and implementation of spin-echo protocol~\cite{levine2018high}. Decoherence caused by spontaneous emission from intermediate state can be reduced by increasing laser power with a larger inter-mediate state detuning, or by employing one photon excitation scheme~\cite{madjarov2020high}. 

\subsection{Scalability of neutral atom system}
\label{subsec:sub3.2}

Scalability constitutes one of the central challenges for implementing large-scale quantum computation. A rough estimate shows the number of physical qubits needed for a rudimentary universal quantum computer easily extends to more than ${10}^{6}$~\cite{saffman2019quantum}. To run a quantum algorithm with an intended circuit depth of ${10}^{10}$, noisy physical qubits with practical error rate at the level of ${10}^{-4}$ would most likely seek help from QEC. With thousands of physical qubits typically needed to make up a logical qubit, quantum algorithm of a few hundred of logical qubits could easily blossom into roughly millions of physical qubits required.

From this perspective, neutral atoms maybe the most promising platform for constructing QEC codes, since only weak magnetic dipole-dipole and vdW interactions exist between ground state atoms, making it possible to closely trap many atoms in large scale with a rich variety of configurations as demonstrated in optical tweezer arrays or magnetic trap arrays~\cite{leung2014magnetic}.

Neutral atoms are inherently identical, thus the requirement for physical resources like laser frequencies etc. do not grow much with the scaling up of qubits. However, neutral atom based systems suffer from cross-talks when single-site addressing is not perfectly satisfied during gate operations or spontaneous emissions from imaging atoms are reabsorbed by nearby atoms during qubit measurements. 

The most challenging issue for neutral atom based QC platform is the stochastic nature of atom loading into individual traps. Collisional blockade~\cite{schlosser2001sub} limits the loading probability of single atom into a small volume trap site to around 50\%. One possible approach to circumvent is to raise the loading probability either by augmenting with blue detuned catalysis light~\cite{lester2015rapid} or using superfluid-Mott insulator transition in optical lattice at the expense of longer experimental cycle time~\cite{weitenberg2011single}, or as demonstrated more recently with immersion cooling based on more advanced control and manipulations of trapping potentials~\cite{yang2020cooling}. However, probability for simultaneous loading of a large number of atoms still becomes exponentially small below a critical level for single atom loading efficiency.

\begin{figure*}
	\centering
	\includegraphics[width=\linewidth]{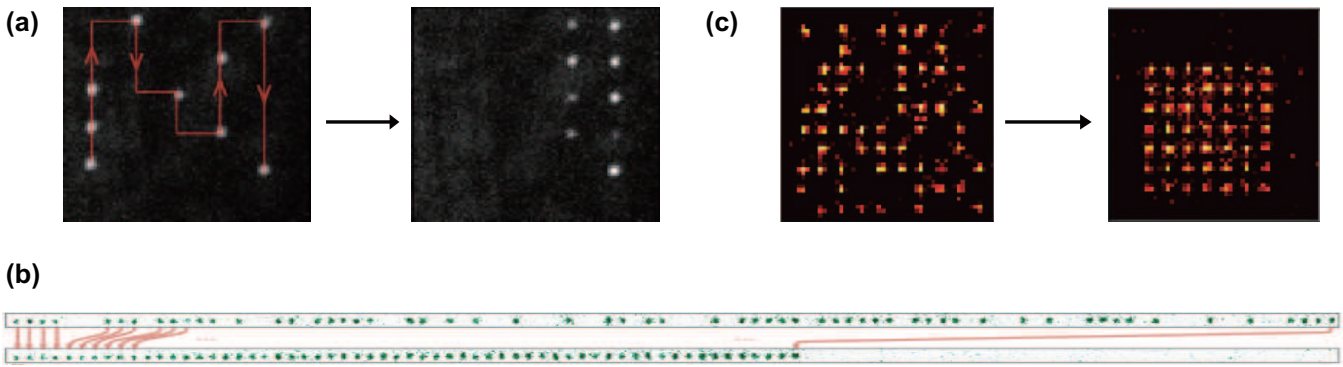}
	\caption{Rearrangement of atoms in dipole trap array. (a) Rearrangement of 9 atoms in a two-dimensional array by varying the holographic optical tweezers formed from a spatial light modulator (adapted from Ref.~\cite{kim2016situ}); (b) Rearrangement of around 50 atoms in a one-dimensional tweeter array formed by 1D acoustic-optical deflector (AOD) through moving the loaded dipole traps to fill in the empty ones (adapted from Ref.~\cite{endres2016atom}); (c) Rearrangement of around 50 atoms in a two-dimensional tweezer array created by a SLM using a moving optical trap generated with a 2D AOD (adapted from Ref.~\cite{barredo2016atom}).}
	\label{fig:fig3_2}
\end{figure*}

A bottom-up approach involves rearranging incompletely filled arrays with moving optical traps (shown in Fig.~\ref{fig:fig3_2}). This approach was independently demonstrated first by several groups in 2016: Kim {\it et al}.~\cite{kim2016situ} prepared 9 atoms in two-dimensional arrays with spatial light modulator; Endres {\it et al}.~\cite{endres2016atom} rearranged 50 atoms in a one-dimensional tweeter array controlled by acoustic-optical deflector (AOD); and Barredo {\it et al}.~\cite{barredo2016atom} operated with a comparable number of atoms in a two dimensional array. In 2018, an arbitrary three dimensional configuration up to 72 atoms was reported~\cite{barredo2018synthetic}. Recently, it is shown that by combining moving optical tweezer steered by a 2D AOD with microtraps generated from a microlens array, a large defect-free cluster with more than 100 atoms can be created \cite{de2019defect}. Sorting atoms sequentially can exponentially increase the probability for defect-free loading. Loss of atom can be mitigated by a similar procedure with lost atoms replenished from a reservoir. 

For atoms in optical lattices with submicron spaced periodic potentials, it is difficult to individually rearrange atoms with micron-scaled moving optical traps. Nevertheless, rearrangement can be carried out using an alternative scheme based on single-site-resolved state flipping and state-sensitive lattice site translation \cite{weiss2004another,vala2005perfect}, which has been demonstrated in one dimensional polarization-synthesized optical lattice \cite{robens2017low} as well as in three dimensional optical lattice \cite{kumar2018sorting} with individual addressing capability.

\subsection{Initialization and detection}
\label{subsec:sub3.3}
Initialization and state detection in a computation cycle most often involves dissipative processes. Initialization of ground state qubits can be simply performed by optical pumping. An additional step for coherent population transfer is needed for qubit encoding into an arbitrary known state.

State measurement is a crucial element for quantum information read-out as well as measurement-based QC or error correction. One simple way to measure the states of atoms is to resonantly push out all atoms of one state and detect the other state by resonant fluorescence imaging of the remaining atoms. This kind of destructive measurement can achieve a high fidelity of 0.9997~\cite{nelson2007imaging}, but cannot discriminate atoms selectively pushed out from the ones lost due to background collision. Moreover, the probability of losing about half the atoms implies one needs to replenish atoms after each measurement, which appreciably increases the cycling time of computation.

To address these issues, {\it in situ} nondestructive measurement techniques based on state selective fluorescence are developed for both optical lattice~\cite{gibbons2011nondestructive} and optical tweezer~\cite{fuhrmanek2011free} systems. Different from nondestructive measurement often referenced to in quantum computation, which categorizes measurements that project qubit state into the same Hilbert space formed by the qubit basis states without excitation into other atom levels, the ``nondestructive'' measurement here emphasizes measurement without atom lost, because atoms exited to other levels can be easily reinitialized with optical pumping. Simultaneous low-loss measurements of multiatoms are realized recently~\cite{martinez2017fast, kwon2017parallel}. Different from destructive measurements, whose errors mainly come from unwanted atoms not blown out or atoms lost from light scattering or background collision during detection, nondestructive measurement incurs additional error from atom heating during imaging. Much efforts have been paid to address trade-offs between more imaging cycles for higher contrast against dark noises from non-ideal photon detectors and from atom losses due to consequent heating. Collection lens with moderately high numerical apertures, deeper optical traps, and careful preparation of polarization states for the probe lights are essential to obtain optimal results.

\begin{figure}
	\centering
	\includegraphics[width=\linewidth]{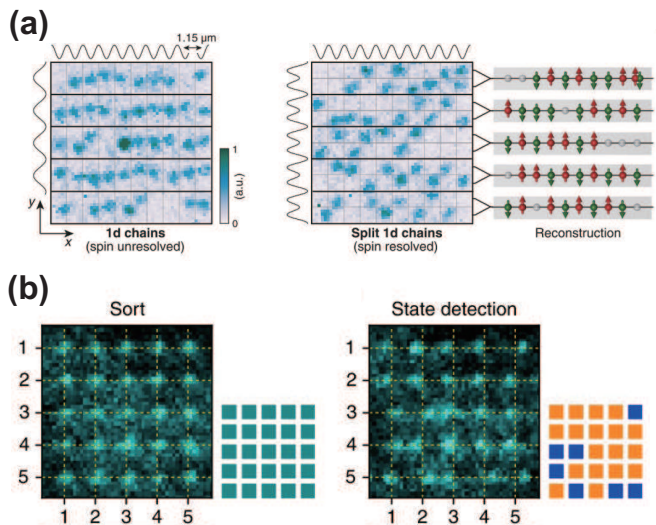}
	\caption{Atomic state detection by spatially resolved imaging. (a) Left: fluorescence image of atoms in five mutually independent one dimensional spin chains. Right: spin-resolved detection. Atoms with different spin states are separated by applying a magnetic field gradient and their wavefunctions projected onto one of the two sites of a double well potential after spatially resolved imaging. (b) Left: fluorescence image of perfect filling for one layer of a 3D optical lattice after sorting a randomly half-filled lattice. Right: atoms with different qubit states are separated by state-dependent optical lattice and their wavefunctions are projected onto one of the two lattice sites after spatially resolved imaging. The dashed yellow grids and square patterns respectively show the atom's initial locations and occupancy maps. Figures (a) and (b) are adapted from Refs.~\cite{boll2016spin}) and \cite{wu2019stern}, respectively.}
	\label{fig:fig3_3}
\end{figure}

In 2016, Boll {\it et al}. resolved atoms in two spin states by shifting them into different locations with a high-gradient magnet field, realizing a detection fidelity of 98\%~\cite{boll2016spin} [shown in Fig.~\ref{fig:fig3_3}(a)]. An improved result is achieved by Wu {\it et al}. in 2019, who employ a state-dependent optical lattice to separate atoms in two qubit states~\cite{wu2019stern} [shown in Fig.~\ref{fig:fig3_3}(b)]. This new measurement scheme inherits the benefits of both former approaches: atom cooling with imaging light and non-destructive state detection. It can be easily extended to reconfigurable dipole-trap arrays.


In order to take the full advantage of scalability by atomic qubits, parallel large-scale qubit measurements present another challenge. Single site imaging at the wavelengh scale of optical lattice~\cite{bakr2009quantum} is carried out with high numerical aperture lens collecting scattered light onto sensitive detectors such as electron multiplying charge coupled device (EMCCD). During post processing of imaging data, Bayesian inference methods are used instead of threshold state discrimination for better state reconstructions in 2D lattices~\cite{martinez2017fast}.

\subsection{Rydberg mediated quantum gates}
\label{subsec:sub3.4}

\subsubsection{Single qubit operations}
As discussed above, atomic qubit states are usually encoded in hyperfine ground state manifolds. As a result, single qubit operations can be performed via microwave or RF field, which can be optimized to give an error rate as small as $\sim10^{-5}$~\cite{sheng2018high}. However, the corresponding transitions preclude straightforward single-site addressing in a large-scale atom array where atoms are closely packed to each other on the correspondingly long wavelength scale. This problem can be overcome by combining the global microwave dressing with a magnet field gradient~\cite{lee2013robust} or a site dependent ac Stark shift induced by a tightly focused laser \cite{xia2015randomized, wang2016single}.

\begin{figure}
	\centering
	\includegraphics[width=\linewidth]{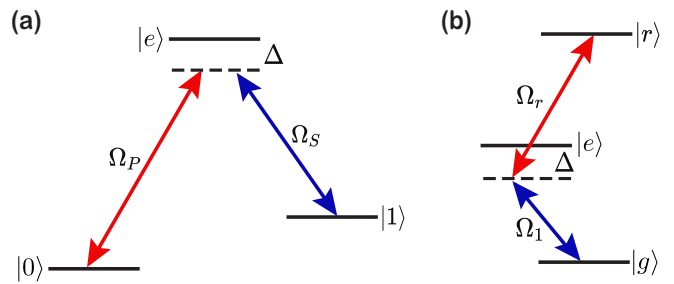}
	\caption{(a) and (b) respectively show the level diagrams for performing single qubit operations in ground state manifolds and with Rydberg state included.}
	\label{fig:fig3_4}
\end{figure}

Single-site addressable one-qubit rotations can also be realized by two focused laser beams using a three-level $\Lambda$-type Raman scheme as shown in Fig.~\ref{fig:fig3_4}(a). In this configuration, coherent transfer between hyperfine ground states $|0\rangle$ and $|1\rangle$ is mediated by an excited intermediate state $|e\rangle$. One way to transfer populations between qubit states $|0\rangle$ and $|1\rangle$ is to use stimulated Raman adiabatic passage (STIRAP) \cite{Bergmann1998}, where the intensity of the coupling beams $\Omega_P(t)$ and $\Omega_S(t)$ are slowly varied in an counter-intuitive time order to keep the system in the dark state $|\Psi_\mathrm{D}\rangle=\cos\Theta|0\rangle-\sin\Theta|1\rangle$ with $\tan\Theta={\Omega_P(t)}/{\Omega_S(t)}$. Alternatively, in the large detuning regime of $\Delta\gg|\Omega_P|,|\Omega_S|$, the intermediate state can be adiabatically eliminated \cite{brion2007adiabatic}, and the resulting dynamics in the rotating frame are governed by an effective Hamiltonian $\hat{H}_\mathrm{eff}=\delta_P|0\rangle\langle0|+\delta_S|1\rangle\langle1|-(\Omega_R|1\rangle\langle0|+\mathrm{H.c.})/2$ with $\delta_\mu=-{\left|\Omega_{\mu}\right|^{2}}/{4 \Delta}, \mu=P,S$ and $\Omega_{R} = {\Omega_{P} \Omega_{S}^{*}}/{(2 \Delta)}$. Such a two-photon Raman transition \cite{yavuz2006fast} enables flexible state manipulation in the ground state manifold, and is also useful for carrying out Rydberg excitation in the $\Theta$-ladder level scheme. As shown in Fig.~\ref{fig:fig3_4}(b), the ground state $|g\rangle$ is coupled to a Rydberg state $|r\rangle$ via an intermediate state $|e\rangle$. Two-photon coupling relaxes more stringent laser-frequency requirement for direct one-photon excitation and facilitates excitations to $nS$ Rydberg states of alkalis which possess nearly isotropic van der Waals interactions.

\subsubsection{Proposals for two-qubit and multi-qubit gates}
Two-qubit quantum gates are among the most important ingredients for universal quantum computation as they can be used to accurately approximate arbitrary unitary operations when combined with certain single-qubit gates \cite{barenco1995elementary}. In neutral atom QC platforms, fast and high-fidelity two-qubit gate operations can be achieved by coupling qubit states to high-lying Rydberg states.

\begin{figure}
	\centering
	\includegraphics[width=\linewidth]{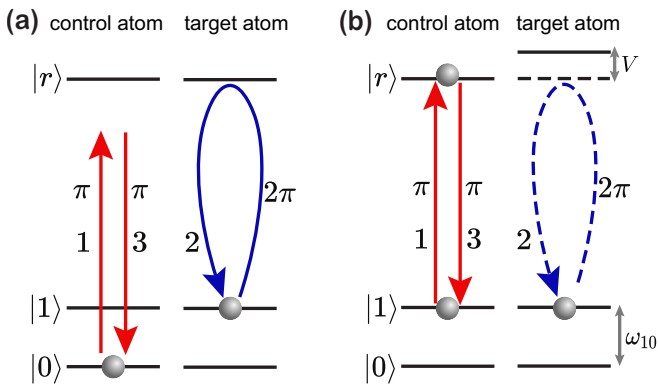}
	\caption{Pulse sequence for the controlled phase gate with Rydberg blockade. (a) and (b) show the dynamics for base state $|01\rangle$ and $|11\rangle$, respectively.}
	\label{fig:fig3_5}
\end{figure}

The pioneering proposal for implementing two-qubit quantum gates with Rydberg atoms is introduced by Jaksch {\it et al}. in 2000 \cite{Jaksch2000}, in which two different regimes for realizing a controlled phase gate (CZ gate) are discussed. In their proposals, one of the qubit states ($|1\rangle$) can be excited to Rydberg state $|r\rangle$ with a Rabi-frequency $\Omega$. When the Rydberg interaction strength $V$ is much smaller than the excitation strength $\Omega$ (weak interaction regime), two atoms can be simultaneously excited into Rydberg state, and their weak interaction contributes a dynamical phase to the corresponding two-qubit state. In the strongly interacting regime $V\gg\Omega$, Rydberg blockade comes into play, and a fast gate operation can be achieved by applying the pulse sequence shown in Fig.~\ref{fig:fig3_5}, which consists of three steps: (1) $\pi$ pulse on the control atom; (2) $2\pi$ pulse on the target atom; (3) $\pi$ pulse on the control atom. If the initial state of the atoms is $|11\rangle$, blockade forbids Rydberg excitation of the target atom due to the large interaction induced level shift $V$. It is easy to verify that the above pulse sequence induces a CZ-gate unitary evolution $\hat{U}_\mathrm{CZ}=2|00\rangle\langle00|-\mathbb{I}$. Unlike the case of weak interaction, the blockade regime is insensitive to qubit positions, which makes it an experimentally favored choice as it is more robustly implemented. By performing $\pi/2$ single-qubit rotations before and after the pulse sequence, the CZ gate is transformed into a controlled NOT (C-NOT) gate.

The long-range feature of Rydberg interaction also facilitates direct construction of multi-qubit gates, in which the conditional dynamics involve multiple control or target qubits. A number of designs are presented, M\"{u}ller {\it et al}. propose a fan-out gate C-NOT$^n$ based on Rydberg blockade and electromagnetically induced transparency (EIT), with which a single control atom switches the states of $n$ target atoms \cite{Muller2009}; Isenhower, Saffman, and M{\o}lmer suggest a multi-qubit Toffoli gate C$_n$-NOT \cite{Isenhower2011}, where spin-flip of the target qubit is conditioned on $n$ control qubits; Su {\it et al}. introduce a multi-qubit controlled phase gate via modified Rydberg antiblockade \cite{su2018one}; Shi outlines a scheme for three-qubit Deutsch gate \cite{Shi2018}. These multi-qubit gates represent highly parrallel quantum operations, which support one-step generation of highly entangled states \cite{gujarati2018rydberg}, simulations of many-body interactions \cite{Weimer2010}, as well as efficient implementations of quantum algorithms \cite{molmer2011efficient}.

In the past few years, tremendous amount of theoretical studies on Rydberg gates lead to the development of many other schemes, aimed at improving the performance of Rydberg mediated quantum gates. One plausible route towards this goal is to go beyond the simple blockade proposal by designing novel pulse sequence such as parallel laser driving with a phase jump \cite{Levine2019,crane2020rydberg}, or by engineering intrinsic interaction mechanisms such as employing Stark-tuned F{\"o}rster resonance \cite{Beterov2016,Beterov2018,Huang2018} and microwave dressing \cite{petrosyan2014binding,shi2017annulled,young2020asymmetric}. Alternatively, by using shaped pulses \cite{Theis2016,Saffman2020} and adiabatic passages \cite{Beterov2013,Rao2014,wu2017rydberg,Petrosyan2017,yu2019adiabatic,Mitra2020,Khazali2020}, one can build up high-fidelity Rydberg gates that are robust against control errors. To shorten the operation time of adiabatic passage while preserving its robustness, improved protocols such as shortcuts to adiabaticity \cite{Shen2019,Liao2019} and non-adiabatic holonomic methods \cite{Zhao2017,Kang2018,Zhao2018} have also found their ways into Rydberg quantum gate designs recently.

\subsubsection{Experimental realization of Rydberg quantum gates}
Two-qubit gate based on the original Rydberg blockade protocol is first realized by Isenhower {\it et al}. in a two-atom configuration \cite{Isenhower2010} and later an improved higher fidelity one by Zhang {\it et al}.~\cite{Zhang2010}. This protocol is soon extended to a 2D qubit array~\cite{Maller2015} and its performance is further refined by using an upgraded setup~\cite{Graham2019}. In such a 2D system where atoms are confined by a grid of repulsive walls and addressed by tightly focused control beams, the CZ gate is effective for creating a Bell state with a fidelity $\mathcal{F}_\mathrm{Bell}=0.86$, with the remaining errors mainly coming from finite atom temperature and laser noise. Similar gate protocol is investigated in a two-atom system consisting of different atomic isotopes~\cite{Zeng2017}, with a prepared Bell state fidelity at $\mathcal{F}_\mathrm{Bell}=0.59$. In such a mixture system, different transition frequencies are employed to individually address either atom, which suppresses cross-talks between different qubits such that stronger Rydberg interaction at shorter interatomic separation distance becomes viable.


Two-qubit gate working in the weak interaction regime with both atoms excited to Rydberg states is also realized. In the experiment reported in Ref.~\cite{Jo2020}, ground-Rydberg entanglement between atoms at a distance larger than Rydberg blockade radius is established with a fidelity $\mathcal{F}_\mathrm{Bell}=0.59$. A similar protocol is demonstrated in a trapped Rydberg ion ($^{88}\mathrm{Sr}^+$) setup \cite{Zhang2020}, which achieves ground state entanglement fidelity $\mathcal{F}_\mathrm{Bell}=0.78$ with a double STIRAP pulse sequence that adiabatically drives the qubit to a Rydberg state and transfers it back to the initial ground state. 

\begin{figure}
	\centering
	\includegraphics[width=\linewidth]{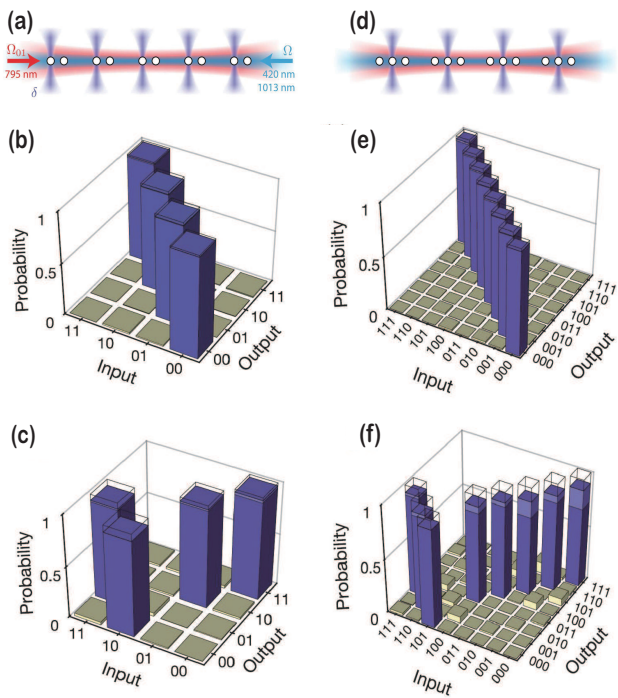}
	\caption{Quantum gates and their implemented performances reported in Ref.~\cite{Levine2019}. (a), (b), and (c) respectively show the configuration, initialization fidelity, and truth table fidelity for two-qubit C-NOT gate. (d), (e), and (f) respectively show the configuration, initialization fidelity, and truth table fidelity for three-qubit Toffoli gate. Figures (a)-(f) are adapted from Ref.~\cite{Levine2019}.}
	\label{fig:fig3_6}
\end{figure}

The increasingly promising potential of Rydberg mediated quantum gates is highlighted by a recent experiment reported in Ref.~\cite{Levine2019}. In this work, a novel approach for implementing two-qubit CZ gate is introduced and demonstrated, with which the fidelity of the converted C-NOT gate is increased to $\mathcal{F}_\mathrm{CNOT}=0.941$ [see Figs.~\ref{fig:fig3_6}(a)-\ref{fig:fig3_6}(c)]. Such a CZ gate relies on a simple and fast pulse sequence to drive nearby atoms within blockade region, consisting of only two global laser pulses with precisely controlled detunings and phase jumps. This protocol can also be extended to build up a multi-qubit gate, such as the three-qubit Toffoli gate with a measured fidelity of $\mathcal{F}_\mathrm{Toff}=0.837$ [see Figs.~\ref{fig:fig3_6}(d)-\ref{fig:fig3_6}(f)].

Based on Rydberg blockade dynamics, atoms can be directly entangled without performing any preset universal two-qubit gate operations. For entanglement between ground state and Rydberg state atoms, several recent experiments report impressively high fidelities, e.g., $\mathcal{F}_\mathrm{Bell}=0.97$ in alkali atom ($^{87}$Rb) system \cite{levine2018high}, $\mathcal{F}_\mathrm{Bell}=0.98$ in alkali earth atom ($^{88}$Sr) system \cite{madjarov2020high}. Once the ground Rydberg entangled state is prepared, it is easy to map the entanglement onto ground state manifold, as demonstrated by Ref.~\cite{Wilk2010} with $\mathcal{F}_\mathrm{Bell}=0.75$ and later improved to $\mathcal{F}_\mathrm{Bell}=0.81$ in~Ref.~\cite{picken2018entanglement}. The ground state entanglement can also be directly produced by employing ground state spin-flip blockade arising from Rydberg dressing induced interaction ($\mathcal{F}_\mathrm{Bell}=0.81$) \cite{jau2016entangling}.

\subsection{Near-term applications}
\label{subsec:sub3.5}
In spite of the steadily increasing number of atomic qubits and the quantum gate fidelities realized in the state-of-the-art experiments, universal quantum computation with QEC remains a long-term goal, considerably out of reach even with scaled up system based on current platforms. However, on the noisy intermediate-scale quantum (NISQ) level, neutral atom based systems can still accomplish several complex computing tasks and exhibit certain quantum features and advantages.

One of the near-term applications for neutral atom based platforms is to solve complex optimization problems by measuring the final states of time evolved quantum many-body systems. A promising paradigm towards this goal is quantum annealing algorithm (QAA), where the solution to a problem is obtained by adiabatically driving the system towards the instantaneous ground state of the objective Hamiltonian. In 2013, Keating {\it et al}. proposed a Rydberg dressing based quantum annealer \cite{keating2013adiabatic} which can be used to solve quadratic unconstrained binary optimization problem. Later on, by utilizing Rydberg-mediated four-body interaction, Glaetzle {\it et al}. introduced a universal quantum annealer with all-to-all couplings \cite{glaetzle2017coherent}. It is then found that anisotropy of the Rydberg interaction can also facilitate the construction of a different universal annealer \cite{qiu2020programmable}. Besides QAA, hybrid quantum-classical variational approaches such as quantum approximate optimization algorithms (QAOA) can also be implemented in Rydberg atom systems. Recently, it is found that 2D neutral atom systems are suitable for solving the maximum independent set problem~\cite{pichler2018quantum} and MaxCut problem~\cite{zhou2020quantum} based on QAOA.

\section{Quantum Simulation with Rydberg Atoms}\label{sec:sec4}
Controllable large-scale quantum systems not only offer the potential to construct universal quantum computers, they are also ideally suited for building up quantum simulators. Quantum simulation aims at using a synthetic quantum system to emulate a real-world many-body physical problem based on a model Hamiltonian, which in general is hard to solve on a classical computer due to exponential growth of the Hilbert space size with increasing particle numbers \cite{feynman1982simulating}. Recent years have witnessed great successes in simulating many-body physics on various platforms, among which neutral atom system with Rydberg interactions appears to be a promising choice. In Rydberg atom system, strong and tunable interactions in combination with available coherent controls and dissipation managements make simulations of a rich variety of many-body problems accessible, such as simulations of coherent spin models, many-body motional dynamics, and driven-dissipative systems that will be discussed later in this section.
\subsection{Coherent spin model}
\label{subsec:sub4.1}
One of the most important research directions for Rydberg atom many-body system is to simulate coherent spin models. To this end, atoms are first loaded into a given lattice configuration (or an ensemble) and prepared in a specific initial state as described in subsections \ref{subsec:sub3.2} and \ref{subsec:sub3.3}. Subsequently, the system evolves under coherent driving in the presence of Rydberg interactions, during which the motional degree of freedom for each atom can be regarded as frozen if the evolution time is sufficiently short. In such a frozen gas limit, the dynamics only alters the internal atomic states, which can effectively emulate an interacting spin system.

The state-dependent interaction in Rydberg atom systems makes it a natural platform to implement analog simulation of diverse spin models in condensed matter physics. For example, Rydberg atom system subjected to vdW interaction can be readily mapped to the quantum Ising model, where each atom constitutes a pseudo-spin $1/2$ system consisting of a ground state $|g\rangle$ and a Rydberg state $|r\rangle$. The corresponding Hamiltonian in the rotating frame reads
\begin{equation}
\hat{H} = \frac{\Omega}{2}\sum_i\hat{\sigma}_{x}^{(i)} + \Delta \sum_i\hat{\sigma}_{rr}^{(i)} + \sum_{i<j}V_{ij}\hat{\sigma}_{rr}^{(i)}\hat{\sigma}_{rr}^{(j)}, \label{eq:eq4.1}
\end{equation}
where $\hat{\sigma}_{\alpha\beta}^{(i)}=|\alpha\rangle_i
\langle \beta|$ and $\hat{\sigma}_{x}^{(i)}=\hat{\sigma}_{rg}^{(i)}+\hat{\sigma}_{gr}^{(i)}$ denote the spin operators for the $i$th atom, $\Omega$ and $\Delta$ respectively denote Rabi frequency and detuning associated with the coherent laser driving, and $V_{ij}=C_6/|{\bm r}_i-{\bm r}_j|^6$ is the vdW interaction between $i$th and $j$th atoms when they are both excited to Rydberg state. By applying the transformation $\hat{\sigma}_{rr}^{(i)}=(\hat{\sigma}_{z}^{(i)}+1)/2$, one can see that Eq.~(\ref{eq:eq4.1}) is equivalent to Ising model in the presence of transverse and longitudinal magnetic fields. Such a model Hamiltonian is widely studied \cite{olmos2009fermionic,pohl2010dynamical,weimer2010two,lesanovsky2011many,basak2018periodically}, and its simulation accuracy has been carefully examined in a recent experiment \cite{de2018accurate}.

\begin{figure}
	\centering
	\includegraphics[width=\linewidth]{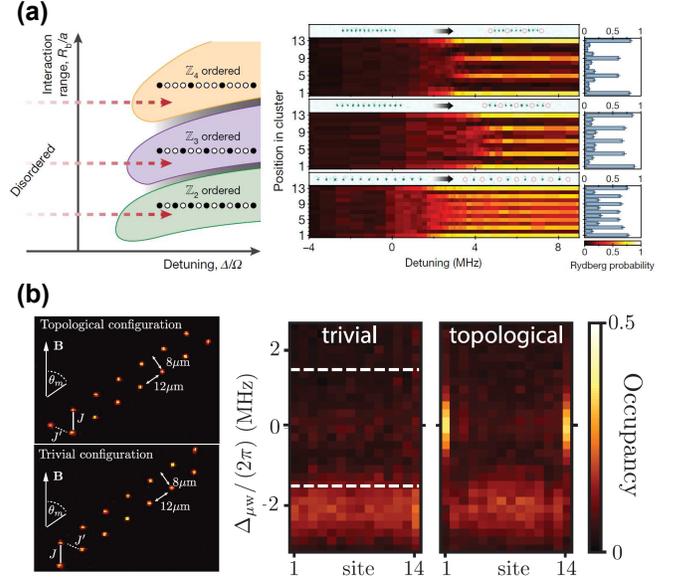}
	\caption{(a) Left figure: ground-state phase diagram of the Rydberg Ising chain [see Eq.~(\ref{eq:eq4.1})]. Right figure: adiabatic preparation of ordered ground states in a 12-atom chain by driving through quantum phase transitions for different values of interaction range $R_b/a$. (b) Left figure: topological and trivial staggered chain configurations for realizing SSH model. Right figure: site-resolved single-particle eigenstate excitation spectrum probed by weak microwave field driving at a detuning $\Delta_{\mu\mathrm{w}}$. Panels (a) and (b) are adapted from Ref.~\cite{bernien2017probing} and Ref.~\cite{deLeseleuc2019observation}, respectively.}
	\label{fig:fig4_1}
\end{figure}

To date, a lot of experiments have attempted to realize the above introduced Ising model, and here we briefly review the impressive works carried out in an ordered lattice configuration. In 2012, Schau{\ss} \emph{et al}. realize the above Ising Hamiltonian in a 2D optical lattice and observe Rydberg blockade induced dynamical crystallization at resonant driving ($\Delta=0$) \cite{schauss2012observation}. Continued studies are reported in Ref.~\cite{schauss2015crystallization}, where the system is instead adiabatically driven to its crystalline ground state. The Ising model is also simulated in optical tweezer arrays. In 2016, Labuhnin \emph{et al}. investigate quench dynamics in a partially filled 2D atom array of arbitrary shape \cite{labuhn2016tunable}. A major breakthrough is achieved by combining defect-free preparation of atom array with Rydberg excitation. Based on this technique, Bernien \emph{et al}. carry out its simulation in a 1D defect-free array with up to 51 atoms \cite{bernien2017probing}. In this experiment, by adiabatically varying the detuning $\Delta(t)$ from positive to negative, they keep the system in the ground state and observe a transition from the disorder phase to a crystalline phase of $Z_2$ order $|rgrg\cdots gr\rangle$, $Z_3$ order $|rggrgg\cdots ggr\rangle$, or $Z_4$ order $|rgggrggg\cdots gggr\rangle$, which is determined by the ratio of blockade radius $R_b$ to lattice constant $a$ [see Fig.~\ref{fig:fig4_1}(a)]. Many other interesting problems are addressed in similar setups, including facilitation and localization dynamics \cite{marcuzzi2017facilitation}, establishment of detailed balance during thermalization \cite{kim2018detailed}, and Kibble-Zurek mechanism in critical dynamics \cite{keesling2019quantum}. Antiferromagnetic correlations are also observed in 2D optical lattice \cite{guardado2018probing} as well as in 2D reconfigurable tweezer array \cite{lienhard2018observing}. Recently, 2D tweezer array system is increased to a remarkably large scale $\sim200$ qubits in various lattice geometries \cite{scholl2020programmable,ebadi2020quantum}.

In addition to direct Rydberg excitation, Rydberg dressing discussed in subsection \ref{subsec:sub2.3} provides an alternative way to implement quantum Ising model. In the dressing protocol, two internal ground states are used to encode spin-up and spin-down states and the Ising interaction takes the form of Eq.~(\ref{eq:eq9}). Coherent many-body dynamics of Ising quantum magnets built up by Rydberg dressing are experimentally studied both in an optical lattice \cite{zeiher2016many,zeiher2017coherent} and in an atomic ensemble \cite{borish2020transverse}.

Resonant dipole-dipole interaction between Rydberg atoms can be used to simulate a new type of many-body spin dynamics. For example, a system containing two dipole coupled Rydberg states $|s\rangle$, $|p\rangle$ can be mapped to a spin-$1/2$ XY model
\begin{equation}
\hat{H} = \sum_{i<j}J_{ij}\left(\hat{\sigma}_{+}^{(i)}\hat{\sigma}_{-}^{(j)}+\hat{\sigma}_{-}^{(i)}\hat{\sigma}_{+}^{(j)}\right), \label{eq:eq4.2}
\end{equation}
where $\hat{\sigma}_{+}^{(i)}=|s\rangle_i\langle p|$ ($\hat{\sigma}_{-}^{(i)}=|p\rangle_i\langle s|$) is raising (lowering) operator for the $i$th spin, and $J_{ij}=C_3(\theta_{ij})/|{\bm r}_i-{\bm r}_j|^3$ is the dipolar interaction with $C_3(\theta_{ij})\propto1-3\cos^2\theta_{ij}$. In this model, Rydberg interaction manifests itself in a coherent excitation transfer between nearby atoms, which is observed in a three-atom system as reported by Ref.~\cite{barredo2015coherent}. Such a long-range XY interaction also leads to nontrivial many-body relaxations as observed in an atomic ensemble \cite{orioli2018relaxation}. By mapping spin excitations to hard-core bosons, this model can be further used to study symmetry protected topological order of strongly interacting bosons, which has been demonstrated in a staggered atom array configuration shown in Fig.~\ref{fig:fig4_1}(b) \cite{deLeseleuc2019observation}. In this work, anisotropy of the interaction facilitates the construction of a Su-Schrieffer-Heeger (SSH) model, which possesses chiral symmetry and can give rise to a single-particle edge state as well as a fourfold ground state degeneracy protected by such a symmetry. Recently, it is shown that dipolar exchange interaction is capable of simulating density dependent Peierls phase \cite{lienhard2020realization}, which establishes an essential step towards observations of many exotic topological phenomena \cite{peter2015topological,weber2018topologically}.

Different from analog simulation that starts from the microscopic Hamiltonian as described above, digital quantum simulation is carried out by decomposing total unitary evolution operator $\hat{U}(t)$ to sequences of discrete unitary gates \cite{georgescu2014quantum}. Applying digital simulation paradigm to Rydberg atom system makes it feasible to explore a broader range of spin dynamics, such as simulation of multiparticle interactions \cite{Weimer2010,weimer2011digital,bohrdt2020multiparticle} and quantum cellular automaton dynamics \cite{lesanovsky2019non,gillman2020nonequilibrium}.

The great success in simulating coherent spin dynamics highlights the potential for Rydberg atom system to implement a practical simulator. In the future, we expect this type of simulation will help us gaining further insights into many-body phenomena of interacting spins, studying spin liquid states \cite{lesanovsky2012liquid} and quantum many-body scars \cite{turner2018weak,ho2019periodic,choi2019emergent,bluvstein2020controlling}, implementing various spin models via Rydberg dressing \cite{glaetzle2015designing,van2015quantum,lan2015emergent}, observing complex orders in higher-dimension lattices \cite{ji2011two,samajdar2020complex}, simulating lattice gauge theory \cite{zhang2018quantum,notarnicola2020real,celi2020emerging,surace2020lattice} and high-energy physics \cite{liu2020realizing}, etc.


\subsection{Atomic and electronic motional dynamics}
\label{subsec:sub4.2}
In the previous subsection, we focus on spin dynamics in the frozen gas limit. For more realistic situation, by taking atomic and electronic center of mass motion into consideration, it is possible to investigate associated physics in a broader scope. In this subsection, we introduce several promising strategies along this direction. We focus on many-body physics here, although Rydberg atom system can also help us to gain insights into few-body physics such as ultralong-range molecule state \cite{li2011homonuclear,shaffer2018ultracold,hollerith2019quantum,fey2020ultralong}.

\begin{figure}
	\centering
	\includegraphics[width=\linewidth]{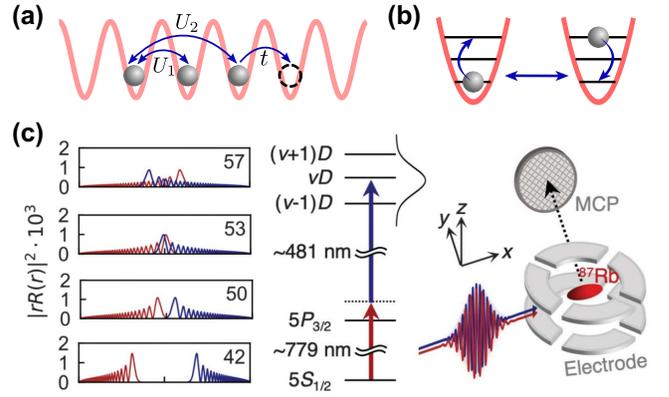}
	\caption{(a) Simulation of Hubbard model with Rydberg dressed ground-state atoms in an optical lattice. Here, $U_1$ and $U_2$ respectively denote the nearest-neighbor and next-nearest-neighbor interactions induced by Rydberg dressing, and $t$ denotes tunneling rate between neighboring sites. (b) Rydberg interaction induced photon exchange between two atoms in deep optical traps. (c) Left: radial wavefunctions of Rydberg electrons in neighboring lattice sites for different principle quantum numbers. Right: Rydberg excitation and ion detection scheme for studying many-body electron dynamics. Figure (c) is adapted from Ref.~\cite{mizoguchi2020ultrafast}.}
	\label{fig:fig4_2}
\end{figure}

First, applying Rydberg dressing into ensemble of ground-state atoms can enrich properties of ultracold quantum gases. For example, when a Bose-Einstein condensate is off-resonantly dressed to a Rydberg state, its wavefunction $\Psi({\bm r})$ evolves according to a modified Gross-Pitaevskii equation
\begin{equation}
i\partial_t\Psi=\left[\frac{-\nabla^2}{2m}+g|\Psi|^2+\int d{\bm r}^\prime U({\bm r}-{\bm r}^\prime)|\Psi({\bm r}^\prime)|^2\right]\Psi,
\end{equation}
where $m$ is the mass of the atom, $g$ is the contact interaction strength, and $U({\bm r}-{\bm r}^\prime)$ denotes Rydberg dressing induced nonlocal interaction strength. Inclusion of such a nonlocal interaction term is essential for the creation of a novel roton-maxon excitation spectrum \cite{henkel2010three} as well as the formation of a supersolid vortex crystal \cite{henkel2012supersolid}. Applying similar dressing scheme into optical lattice can facilitate quantum simulation of the Hubbard model with long-range soft-core interactions [see Fig.~\ref{fig:fig4_2}(a)]. Theoretical studies reveal rich exotic phases of matter emerging from such an extended Hubbard model, such as topological Mott insulators \cite{dauphin2012rydberg}, cluster Luttinger liquids \cite{mattioli2013cluster}, superglass phases \cite{angelone2016superglass}, and supersolidities \cite{li2018supersolidity,zhou2020quench}. For atoms subjected to a synthetic gauge field, Rydberg dressing scheme is capable of simulating fractional quantum Hall phases of bosons \cite{grass2018fractional}.

When an atom is deeply trapped in a potential, it cannot hop to nearby lattice sites. In this case, distance-dependent Rydberg interaction can modify the vibrational state of an atom in the trap, giving rise to effective phonon dynamics, as shown in Fig.~\ref{fig:fig4_2}(b). Coupling between vibrational and internal degrees of freedom can result in complex excitation dynamics (e.g., dissipative blockade effect studied by Ref.~\cite{li2013nonadiabatic}) and provides an approach to generating nonclassical motional states \cite{buchmann2017creation}. Recently, it is found that a state-insensitive phonon swap induced by Rydberg dressed interaction is useful for sympathetic cooling of an atomic quantum simulator \cite{belyansky2019nondestructive}. The interplay between spin and vibrational dynamics can also be used to engineer polaronic quasi-particle excitations \cite{mazza2020vibrational} and three-body interactions \cite{gambetta2020engineering}. Motional dynamics of a flexible array of Rydberg atoms without external trapping potential is intriguing as well, which for instance can simulate a Newton-cradle-like excitation transport \cite{wuster2010newton}.

Excitations of atoms to their Rydberg states via ultrashort coherent laser pulses provides an approach to exploring many-body physics with strongly interacting electrons. In such an ultrafast quantum simulator, many Rydberg electrons are excited to orbits far from their cation cores. As a result, wavefunctions of these electrons spatially overlap with each other, which then generates strong many-body correlations induced by Coulomb interactions. Based on this idea, Takei \emph{et al}. observe coherent many-body electron dynamics in an atomic ensemble by time-domain Ramsey interferometry \cite{takei2016direct}. Subsequently, Mizoguchi \emph{et al}. study many-electron dynamics with an atomic Mott insulator in an optical lattice \cite{mizoguchi2020ultrafast}. As observed in their experiment [see Fig.~\ref{fig:fig4_2}(c)], overlap of nearby Rydberg electron wavefunctions leads to drastic change of ion-counting statistics and sharp increase of avalanche ionization, which opens up a way to investigate metal-like phases of Rydberg gases.

\subsection{Dissipative many-body physics}
\label{subsec:sub4.3}
In a realistic Rydberg atom system, coherent driving offered by external field often competes with dissipation induced by coupling with environment. Such a controllable driven-dissipative system with strong and nonlocal Rydberg-Rydberg interactions can be used to simulate many-body phenomena distinct from their fully coherent counterparts, e.g., dynamical phase transitions that are far from equilibrium \cite{henkel2008non}. 

Evolution of such an open many-body system is often governed by the master equation $\partial_t\hat{\rho}=-i[\hat{H},\hat{\rho}]+\hat{\mathcal{L}}\hat{\rho}$, where $\hat{\rho}$ is the density matrix of the system, $\hat{H}$ is the Hamiltonian describing coherent driving and interaction, and $\hat{\mathcal{L}}$ denotes the Liouvillian operator associated with dissipation. For a simple two-level system with Hamiltonian given by Eq.~(\ref{eq:eq4.1}), spontaneous decay of Rydberg states can be described by the Liouvillian term $\hat{\mathcal{L}}\hat{\rho}=\Gamma\sum_n[\hat{\sigma}_{gr}^{(n)}\hat{\rho}\hat{\sigma}_{rg}^{(n)}-\{\hat{\sigma}_{rr}^{(n)},\hat{\rho}\}/2]$. Such a depopulation term competes with coherent excitation, which together with Rydberg interaction induced blockade or antiblockade can give rise to highly correlated steady state. Theoretical studies of this model reveal a host of interesting steady-state properties, such as non-equilibrium phase transitions \cite{lee2011antiferromagnetic,ates2012dynamical,hu2013spatial,qian2015dynamical}, steady-state crystallization and aggregation \cite{honing2013steady,hoening2014antiferromagnetic,schonleber2014coherent,weimer2015variational,malossi2014full}, as well as bistabilities and metastabilities \cite{sibalic2016driven,letscher2017bistability}. Experimental observations of the associated optical bistability and hysteresis in this system are reported in Refs.~\cite{carr2013nonequilibrium,demelo2016intrinsic,weller2016charge,ding2020phase}.
\begin{figure}
	\centering
	\includegraphics[width=\linewidth]{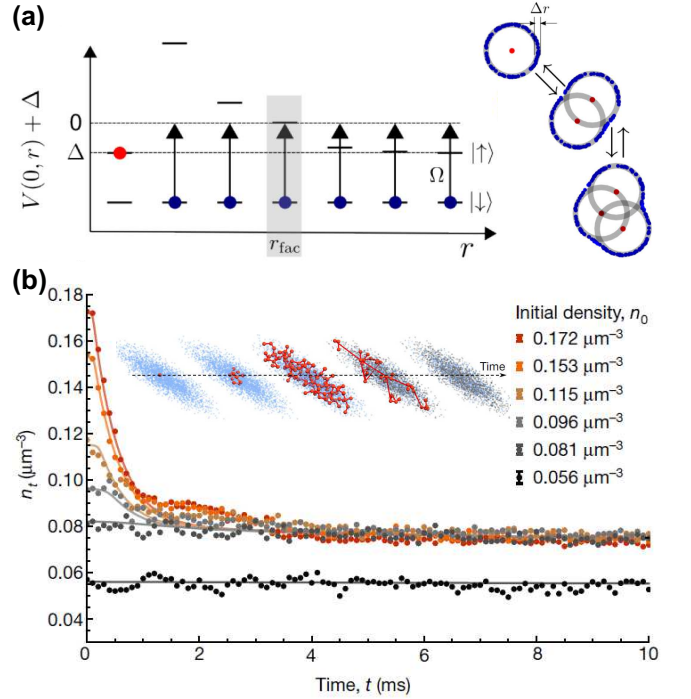}
	\caption{(a) Illustration of facilitation dynamics (adapted from Ref.~\cite{lesanovsky2014out}), where $V(0,r)$ in the left panel denotes vdW interaction between a Rydberg atom located at the origin and another Rydberg atom located at a distance $r$ from the origin. The right panel shows facilitation induced aggregation. (b) Evolutions of the remaining atom density in a driven-dissipative Rydberg gas for different initial densities above and below the threshold value (adapted from Ref.~\cite{helmrich2020signatures}). Above a threshold value, the remaining total atom density is attracted to the same steady-state value, while for initial densities below the threshold density the dynamics becomes stationary. The inset depicts the corresponding self-organization process, where blue dots and red spheres respectively represent atoms in the ground state and Rydberg state.}
	\label{fig:fig4_3}
\end{figure}

Another important type of dissipation is dephasing of Rydberg states induced by thermal motion of atoms or phase noise of the driving laser, whose corresponding Liouvillian term takes the form $\hat{\mathcal{L}}\hat{\rho}=\gamma\sum_n[\hat{\sigma}_{rr}^{(n)}\hat{\rho}\hat{\sigma}_{rr}^{(n)}-\{\hat{\sigma}_{rr}^{(n)},\hat{\rho}\}/2]$. Such a pure dephasing usually leads to an uncorrelated steady state, but the relaxation process can be highly nontrivial. For example, in the strong dephasing regime, Rydberg interactions result in kinetically constrained dynamics closely related to soft-matter systems \cite{lesanovsky2013kinetic}, where atomic growth rate for Rydberg excitation is strongly affected by the state of its neighboring atoms. In particular, an already excited atom can facilitate Rydberg excitation of another atom located near the facilitation radius $r_\mathrm{fac}$ where the vdW interaction $V$ satisfies $\Delta+V=0$ \cite{lesanovsky2014out,gribben2018quench}. Such a condition can result in correlated growth of Rydberg aggregates [see Fig.~\ref{fig:fig4_3}(a)], whose features are observed already \cite{urvoy2015strongly,valado2016experimental,simonelli2016seeded,bai2020distinct}. As for systems dominated by Rydberg mediated spin-exchange interactions, dephasing can strongly influence the dynamics of excitation transport, which usually exhibits a crossover behavior from quantum ballistic regime to classical diffusive regime \cite{gunter2013observing,schonleber2015quantum,schempp2015correlated,yang2019quantum,whitlock2019diffusive}.

In more realistic cases, both decay and dephasing come into play and a Rydberg excitation can decay to another ground state refrained from being excited again. The cooperation of these processes usually leads to complicated dynamical evolutions and nontrivial steady states, such as epidemic dynamics mimicking the spreading of a disease \cite{perez2017epidemic}, mobile excitation holes in Rydberg superatoms \cite{letscher2017manybody}, and directed percolation (DP) phase transitions \cite{marcuzzi2015non,marcuzzi2016absorbing}. As an important example, DP phase transition behaves as a non-equilibrium phase transition from an absorbing phase with vanishing Rydberg populations to an active phase of finite Rydberg densities and large fluctuations [see Fig.~\ref{fig:fig4_3}(b)]. Experimental signatures of such a phase transition are observed in several setups \cite{gutierrez2017experimental,helmrich2018uncovering,helmrich2020signatures}, which points to a realistic physical system for testing self-organized criticality.

\section{Quantum Optics with Rydberg Atoms}\label{sec:sec5}
In Secs.~\ref{sec:sec3} and \ref{sec:sec4}, we focus on quantum information processing (QIP) with neutral atoms as physical qubits. In a companion direction, interfacing Rydberg atoms with photons opens up a route to study photonic quantum computation and quantum simulation. As flying qubits, individual photons can be easily manipulated and transported over long distances, which makes them natural information carriers for quantum communication and quantum networks \cite{kimble2008quantum}.
\begin{figure*}
	\centering
	\includegraphics[width=0.9\linewidth]{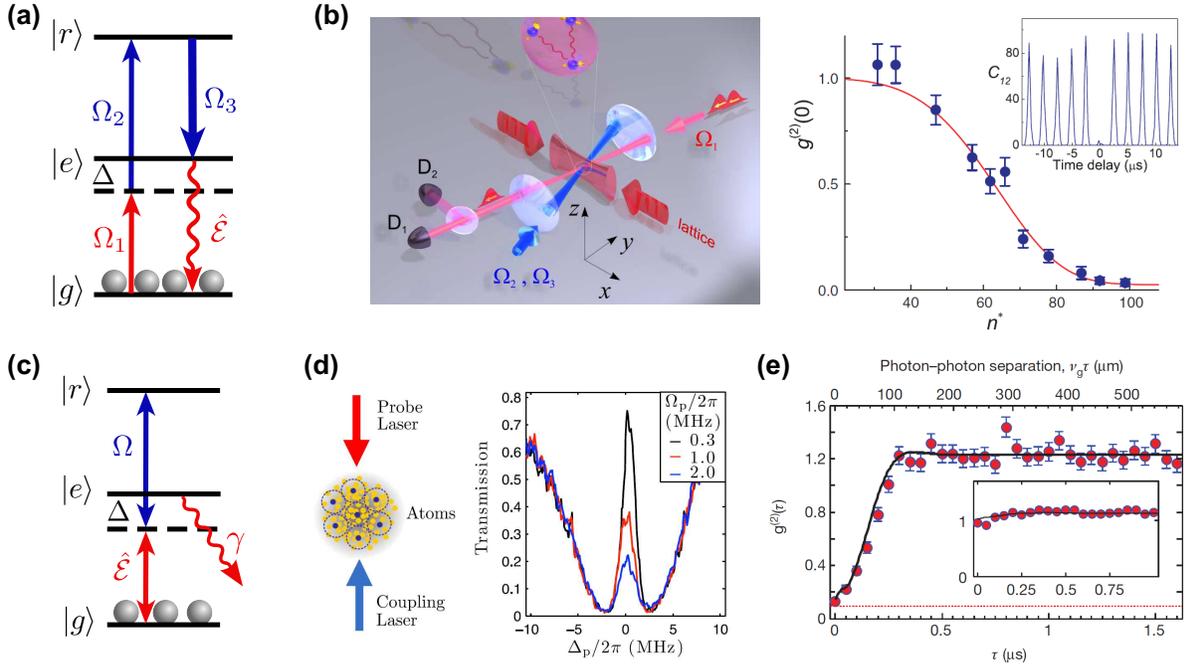}
	\caption{(a) Level diagram of the four-wave mixing scheme for single-photon generation. (b) Experimental setup and measured two-photon correlation functions in Ref.~\cite{dudin2012strongly}. (c) Level scheme of the Rydberg EIT. (d) Schematic of the Rydberg EIT configuration and the measured probe photon transmissions for different intensities (adapted from Ref.~\cite{pritchard2010cooperative}). (e) Two-photon correlation functions for transmitted probe photons in a resonant Rydberg-EIT configuration (adapted from Ref.~\cite{peyronel2012quantum}).}
	\label{fig:fig5_1}
\end{figure*}

The lack of tangible interaction between photons in conventional photonic systems makes it difficult to generate entanglement between photons. Although projective measurement opens up a probabilistic approach of inducing an effective interaction in linear optical setups \cite{kok2007linear}, it would be highly desirable if direct and strong interactions between single photons can be established \cite{chang2014quantum}. Single photons propagating inside an atomic ensemble can be coupled to collective excitations of atoms. If the collective exciation is associated with a Rydberg state $|r\rangle$, it is possible to map the strong interaction between Rydberg atoms onto single photons \cite{firstenberg2016nonlinear,murray2016quantum}. For $N$ atoms initially in the ground state $|g\rangle$, the polariton formed by a collective Rydberg excitation $\sum_n c_n|g_1\cdots r_n\cdots g_N\rangle$ can be described by a continuous bosonic field operator $\hat{\mathcal{S}}({\bm r})$ satisfying $[\hat{\mathcal{S}}({\bm r}),\hat{\mathcal{S}}^\dagger({\bm r}^\prime)]=\delta({\bm r}-{\bm r}^\prime)$ \cite{murray2016quantum}. Considering a simple vdW interaction between atoms in Rydberg state $|r\rangle$, the nonlocal two-body interaction between Rydberg polaritons are given by
\begin{equation}
\hat{V} = \frac{1}{2}\iint \mathrm{d}{\bm r}\mathrm{d}{\bm r}^\prime V({\bm r}-{\bm r}^\prime)\hat{\mathcal{S}}^\dagger({\bm r}^\prime)\hat{\mathcal{S}}^\dagger({\bm r})\hat{\mathcal{S}}({\bm r})\hat{\mathcal{S}}({\bm r}^\prime),\label{eq:eq5.1}
\end{equation}
where $V({\bm r}-{\bm r}^\prime)=C_6/|{\bm r}-{\bm r}^\prime|^6$ denotes vdW interaction between Rydberg excitations located at ${\bm r}$ and ${\bm r}^\prime$. When slowly-varying photonic field $\hat{\mathcal{E}}({\bm r})$ is coupled to the Rydberg polariton $\hat{\mathcal{S}}({\bm r})$, vdW interaction can give rise to strong interaction between photons. In this section, we briefly review recent progresses in quantum optics based on Rydberg atomic ensembles.

\subsection{Single-photon generation and atom-photon entanglement}
\label{subsec:sub5.1}
An efficient single-photon source is indispensable for photonic QIP. While it can be realized by coupling a single emitter (e.g., a quantum dot) with a highly localized optical mode (e.g., a micro-cavity mode), Rydberg atomic ensemble provides an alternative way that works in free space.

The original proposal [see Fig.~\ref{fig:fig5_1}(a)] for Rydberg mediated single-photon generation is based on Rydberg blockade and cooperative spontaneous emission \cite{saffman2002creating}. As a first step of this proposal, excitation beams $\Omega_1$ and $\Omega_2$ induce an effective driving $\hat{H}_\mathrm{d} = \int \mathrm{d}{\bm r}\sqrt{\rho({\bm r})}[\Omega_\mathrm{eff}({\bm r})\hat{\mathcal{S}}^\dagger({\bm r})+\mathrm{H.c.}]$ with $\rho({\bm r})$ the atomic density and $\Omega_\mathrm{eff}({\bm r})$ the effective Rabi frequency. If atoms addressed by the effective coupling $\Omega_\mathrm{eff}({\bm r})$ are within the blockade radius, the strong interaction $\hat{V}$ [Eq.~(\ref{eq:eq5.1})] only allows for the excitation of a single Rydberg polariton $|R\rangle\propto \int\mathrm{d}{\bm r}\sqrt{\rho({\bm r})}\Omega_\mathrm{eff}({\bm r})\hat{\mathcal{S}}^\dagger({\bm r})|G\rangle$, where $|G\rangle=|g_1\cdots g_N\rangle$ denotes the spin-wave vacuum state with all atoms in the ground state $|g\rangle$. Subsequently, a coupling beam $\Omega_3$ transfers the created single Rydberg polariton to an excited state polariton, which can then be efficiently converted into a directed photon in a well-defined mode $\hat{\mathcal{E}}$ based on cooperative spontaneous emission \cite{scully2006directed}.

The above scheme is realized by Dudin and Kuzmich \cite{dudin2012strongly}. In their experiment, a strong antibunching of retrived photons is observed as the principal quantum number $n$ is increased [see Fig.~\ref{fig:fig5_1}(b)], indicative of successful generation of single photons. In addition to Rydberg blockade, they find that interaction induced dephasing can also account for the observed antibunching phenomenon. The high single-photon generation efficiency of this system attracts broad attention. In 2016, Li and Kuzmich combine Rydberg blockade with ground-state storage, significantly increasing the total storage time of the system \cite{li2016quantum}. Li {\it et al}. employ a similar approach to create two orthogonal spin-wave excitations and observe Hong-Ou-Mandel interference between them \cite{li2016hong}. The single-photon generation schemes are also studied in room-temprature atomic gases \cite{ripka2018room} and in steadily improved setups \cite{ornelas2020demand}. The reported single-photon emission efficiency in current experiments remains limited due to the small optical depth within the excitation blockade radius. Recently, Petrosyan and M{\o}lmer propose a scheme to map single-atom excitation to a single photon with dipolar exchange interactions \cite{petrosyan2018deterministic}, which relaxes the need for Rydberg blockade and therefore permits a delocalized excitation that can be retrieved more efficiently.

Besides serving as single-photon source, Rydberg atomic ensemble can also be used to create atom-photon entanglement, which plays a pivotal role in the envisaged quantum internet \cite{kimble2008quantum}. Once single Rydberg polariton is prepared, one can map the state $|R\rangle|0\rangle$ to an atom-light entangled state $(|R\rangle|0\rangle+|G\rangle|1\rangle)/\sqrt{2}$ through a ``half'' retrieval operation, where $|0\rangle$ and $|1\rangle$ denote the Fock states for photons. In 2013, Li, Dudin, and Kuzmich realize such entanglement and indeed observe violation of Bell-inequality \cite{li2013entanglement}. The entanglement in the above protocol is encoded in photonic Fock space, while a polarization entangled scheme usually offers a comparatively better performance in a quantum repeater, as it tolerates a larger propagation phase instability and relaxes the need for photon-number-resolved detectors \cite{sangouard2011quantum}. In 2019, Li {\it et al}. report a novel scheme to establish atom-photon entanglement in the polarization basis \cite{li2019semideterministic}. Quite recently, Yang, Liu, and You propose a scheme to realize atom-photon spin-exchange interaction via Rydberg dressing \cite{yang2020atom}, with which the desired atom-photon entanglement can be robustly established without storage or retrieval of photons.

\subsection{Single-photon quantum-logic operation}
\label{subsec:sub5.2}
Controlling photons by photons requires a nonlinear medium. The nonlinearities of conventional materials at the level of a few photons are relatively small, so that a large number of photons are usually required to observe nonlinear optical effects. Quantum nonlinear optics open the door to studying the regime where a single photon can significantly affect the state of another photon \cite{chang2014quantum}. Such a strong nonlinearity is key to all-optical quantum logic operations.

As explained in the beginning of this section, coupling photonic field with Rydberg polariton endows photons with strong nonlinearities. An efficient way to achieve this coupling is to employ electromagnetically induced transparency (EIT) as shown in Fig.~\ref{fig:fig5_1}(c). In a Rydberg EIT system, the quantized photonic field $\hat{\mathcal{E}}({\bm r})$ and the Rydberg polariton field $\hat{\mathcal{S}}({\bm r})$ form a dark state polariton (DSP) $\hat{\Psi}({\bm r})=\cos(\theta)\hat{\mathcal{E}}({\bm r})-\sin(\theta)\hat{\mathcal{S}}({\bm r})$ \cite{fleischhauer2000dark}. As a result, the strong interaction $\hat{V}$ in Eq.~(\ref{eq:eq5.1}) for the Rydberg component $\hat{\mathcal{S}}({\bm r})$ will induce a nonlocal nonliearity for the photonic field $\hat{\mathcal{E}}({\bm r})$. The original Rydberg-EIT proposal considers weakly interacting regime \cite{friedler2005long}, where two DSPs passing through each other acquire a global phase. In the strongly interacting regime, the nonlinearity can be attributed to a EIT blockade mechanism \cite{ates2011electromagnetically,gorshkov2011photon,petrosyan2011electromagnetically,sevinccli2011nonlocal,liu2014electromagnetically}: the energy of the Rydberg level $|r\rangle$ is modified by the interaction $\hat{V}$, which can destroy the EIT condition. Specifically, when an atom located at ${\bm r}$ is excited to the Rydberg state, the level $|r\rangle$ for an atom at ${\bm r}^\prime$ will experience a shift $V({\bm r}-{\bm r}^\prime)$, which effectively changes the atom at ${\bm r}^\prime$ to a two-level system if $V({\bm r}-{\bm r}^\prime)$ is comparable to or larger than the EIT linewidth $\gamma_\mathrm{EIT}=\Omega^2/|\gamma-i\Delta|$. Unlike a EIT process with vanishing susceptibility, transmission of the incoming light can be significantly modified by such an effective two-level system. For $\gamma\gg\Delta$, the nonlinearity takes a dissipative feature: photon transmission within the radius $z_b=(C_6\gamma/\Omega^2)$ of Rydberg excitation is strongly suppressed. In the regime of $\Delta\gg\gamma$, photons interact with each other via a dispersive nonlocal interaction over the range $z_b=(C_6|\Delta|/\Omega^2)$ \cite{gorshkov2011photon}.

In 2010, Pritchard {\it et al}. demonstrate dissipative nonlinearity of a Rydberg EIT system \cite{pritchard2010cooperative}. Figure \ref{fig:fig5_1}(d) shows their experimental results, where a clear suppression of transmission at the EIT peak appears as the input intensity for the probe field $\hat{\mathcal{E}}$ is increased. The dispersive nonlinearity is then observed by putting Rydberg atoms in a low-finesse cavity and measuring the interaction induced resonance frequency shift \cite{parigi2012observation}. Although these pioneering experiments are performed in the classical regime, they demonstrate the underline physics and open up possibilities to explore quantum nonlinear optics in Rydberg EIT systems.
\begin{figure}
	\centering
	\includegraphics[width=\linewidth]{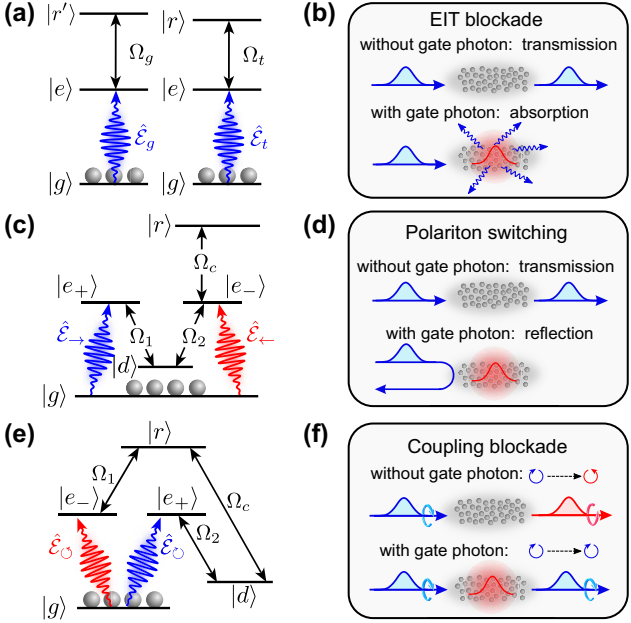}
	\caption{(a) and (b) illustrate level scheme and working mechanism for EIT blockade based switching. (c) and (d) illustrate the target-photon level diagram and the mechanism for polariton switching scheme. (e) and (f) illustrate the target-photon level diagram and the mechanism for coupling blockade scheme. The level schemes for gate photon in polariton switching and coupling blockade protocols are the same as in EIT blockade protocol.}
	\label{fig:fig5_2}
\end{figure}

The quantum regime of Rydberg-EIT nonlinearity is first observed by Peyronel {\it et al}. in 2012 \cite{peyronel2012quantum}. In the experiment, by tightly focusing a probe beam into a waist smaller than the blockade radius $z_b$, input photons are regulated into a train of single photons, as verified by the suppression of the two-photon correlation function displayed in Fig.~\ref{fig:fig5_1}(e). Dispersive quantum nonlinearity is then realized in a similar setup \cite{firstenberg2013attractive} and characterized by an attractive interaction between photons and the appearance of an interaction induced phase accumulation.

An important class of quantum logic operation---single-photon switch or transistor, allows for the control of target-photon transmission with a single gate photon \cite{murray2016many}. Figures \ref{fig:fig5_2}(a) and \ref{fig:fig5_2}(b) display the switching scheme in a Rydberg EIT system. The gate photon in mode $\hat{\mathcal{E}}_g$ is first stored in an atomic ensemble as a collective excitation in Rydberg state $|r^\prime\rangle$. Target photons in mode $\hat{\mathcal{E}}_t$ involving another Rydberg state $|r\rangle$ are then sent into the atomic ensemble. The transmission of the target photon can be suppressed by the stored gate photon, and a large extinction radio can be obtained when the optical depth $d_b$ within the blockade radius is sufficiently high. This scheme is realized by Baur {\it et al}. \cite{baur2014single} in 2014. If the switch can offer a gain, i.e., the number of target photons switched by a single gate photon is above unity, it can work as a single-photon transistor. In 2014, Tiarks {\it et al}. \cite{tiarks2014single} and Gorniaczyk {\it et al}. \cite{gorniaczyk2014single} independently demonstrate this type of single-photon transistor with a gain above 10. The gain of the transistor can be further boosted by tuning near to a F\"{o}ster resonance \cite{gorniaczyk2016enhancement}.

In 2016, Murray, Gorshkov, and Pohl \cite{murray2016many} show there exists certain fundamental limitations to the above introduced optical switch when retrieval of the gate photon is considered. Furthermore, irreversible photon scattering in this scheme prevents the realization of a quantum switch. To address these issues, new schemes are proposed to realize coherent switch. In 2017, Murray and Pohl \cite{murray2017coherent} propose a novel polariton switching mechanism. In this proposal [see Figs.~\ref{fig:fig5_2}(c) and \ref{fig:fig5_2}(d)], a coherent router-type interaction is established, where the target photon in mode $\hat{\mathcal{E}}_\rightarrow$ is reflected to mode $\hat{\mathcal{E}}_\leftarrow$ by a stored gate photon. In 2019, Yang, Liu, and You \cite{yang2019manipulating} introduced a multi-transverse mode scheme, where coherent control of photonic states is mediated by a coupling blockade mechanism [see Figs.~\ref{fig:fig5_2}(e) and \ref{fig:fig5_2}(f)]: the gate excitation can switch the target-photon state by blockading the coupling between modes $\hat{\mathcal{E}}_\circlearrowleft$ and $\hat{\mathcal{E}}_\circlearrowright$.
\begin{figure}
	\centering
	\includegraphics[width=\linewidth]{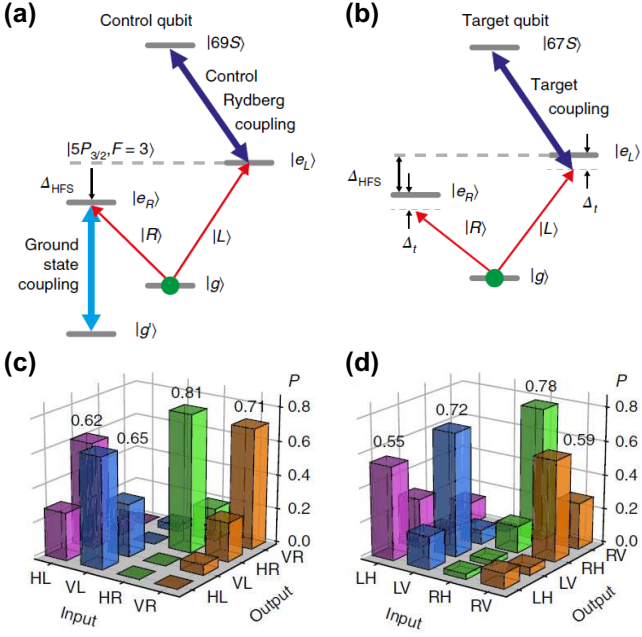}
	\caption{(a) and (b) respectively show level schemes of control and target qubits for realizing a photonic quantum gate. When both photons are in the left circularly polarized state $|L\rangle$, the total wavefunction will pick up an interaction induced phase. (c) and (d) show the measured truth tables of the C-NOT gate. Figures (a)-(d) are adapted from Ref.~\cite{tiarks2019photon}.}
	\label{fig:fig5_3}
\end{figure}

Deterministic quantum-logic gates for photons can also be realized in Rydberg EIT system. In the dispersive regime $\Delta\gg\gamma$, two photons passing through each other will accumulate a global phase $\phi$ to the two-particle wavefunction, which can be used to build up a photonic CZ gate by tuning $\phi=\pi$. Such an interaction induced phase is first measured by Firstenberg {\it et al}. \cite{firstenberg2013attractive}, where $\phi\sim1~\mathrm{rad}$ is reported. A significant phase $\phi\sim\pi$ is then realized by Tiarks {\it et al}. in 2016 \cite{tiarks2016optical}. For a complete logic gate, photons need to be encoded in different qubit states, as in the work of Tiarks {\it et al}. \cite{tiarks2019photon}, where by encoding quantum information in the polarization basis [see Figs.~\ref{fig:fig5_3}(a) and \ref{fig:fig5_3}(b)], they demonstrate a photonic C-NOT gate with a fidelity of $\mathcal{F}_\mathrm{CNOT}=70\%$ [see Figs.~\ref{fig:fig5_3}(c) and \ref{fig:fig5_3}(d)]. There exists several alternative Rydberg-based schemes for realizing a photonic quantum gate, e.g., storing photons in atomic ensembles and utilizing Rydberg interactions between stored spin-wave modes \cite{paredes2014all,khazali2015photon}, employing cavities to enhance atom-photon couplings \cite{das2016photonic,wade2016single,lahad2017induced,sun2018analysis}, or using dipolar exchange interaction together with a symmetry-protected mechanism to construct a robust $\pi$-phase gate \cite{thompson2017symmetry,khazali2019polariton}.

\subsection{Few-body and many-body physics of interacting photons}
\label{subsec:sub5.3}
Rydberg mediated photonic interaction not only supports the construction of quantum-logic operations, it also facilitates quantum simulation with photons. 

\begin{figure}
	\centering
	\includegraphics[width=\linewidth]{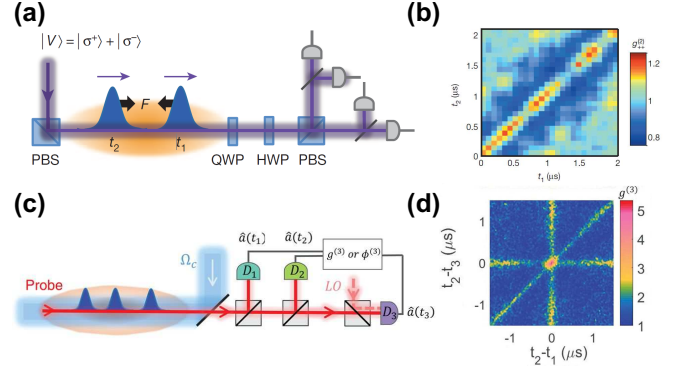}
	\caption{(a) and (b) show the experimental setup for observing two-photon bound states and the measured two-photon correlation function, respectively (adapted from Ref.~\cite{firstenberg2013attractive}). (c) and (d) show the experimental setup for observing three-photon bound states and the measured three-photon correlation function, respectively (adapted from Ref.~\cite{liang2018observation}).}
	\label{fig:fig5_4}
\end{figure}

In fact, simulating few-body dynamics with interacting photons has attracted wide attentions \cite{firstenberg2013attractive,bienias2014scattering,maghrebi2015coulomb,jachymski2016three,gullans2017efimov,busche2017contactless,liang2018observation,stiesdal2018observation,bienias2020exotic,cantu2020repulsive}. Particular interests are paid to photonic molecules---bound states formed by a few photons, which resemble molecule states formed by massive particles subjected to an attractive interaction for bonding. In an atomic ensemble, Rydberg EIT with a large intermediate state detuning $\Delta$ endows a photon with an attraction as well as an effective mass, which opens the door for the existence of photonic bound state. In 2013, Firstenberg \cite{firstenberg2013attractive} observe two-photon bound state in a Rydberg EIT system [see Fig.~\ref{fig:fig5_4}(a)], backed by strong bunching of the two-photon correlation function shown in Fig.~\ref{fig:fig5_4}(b). Subsequently, Bienias {\it et al}. examine scattering resonance in the same system \cite{bienias2014scattering}, and Maghrebi {\it et al}. study a diatomic type photon molecule \cite{maghrebi2015coulomb} formed by a Coulomb-like potential. Three-photon bound state is subsequently studied in Refs.~\cite{jachymski2016three,gullans2016effective} which includes three-body interactions, and in Ref.~\cite{gullans2017efimov} which discusses a photonic version of Efimov trimer. The associated experiment is reported by Liang {\it et al}. \cite{liang2018observation}, in which bunching of the third-order correlation function shown in Fig.~\ref{fig:fig5_4}(b) affirms the existence of three-photon clustering. Recently, Bienias {\it et al}. find that Lennard-Jones-like potentials can be created by tuning near to a F\"{o}rster resonance in a Rydberg EIT configuration \cite{bienias2020exotic}, with which many other exotic few-photon bound states can be observed.

\begin{figure}
	\centering
	\includegraphics[width=0.9\linewidth]{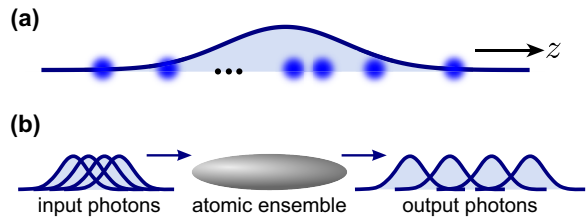}
	\caption{(a) Schematic of a multiphoton pulse propagating along $z$-direction. (b) Illustration of crystallization of photons passing through a Rydberg atomic ensemble.}
	\label{fig:fig5_5}
\end{figure}

Quantum many-body physics lies at the heart of quantum simulation. Many-body physics of interacting photons are particularly interesting because the delocalized nature of a photon can give rise to phenomenon without counterparts in other platforms [see Fig.~\ref{fig:fig5_5}(a)]. As a prominent example, it is predicted that photons passing through a Rydberg atomic ensemble can form a moving-frame crystal \cite{otterbach2013wigner,moos2015many}, as shown in Fig.~\ref{fig:fig5_5}(b). Several experiments exploring this multiphoton scattering regime are carried out, including interaction between many photons and a Rydberg superatom \cite{paris2017free}, single-photon subtraction mediated by many-body decoherence \cite{murray2018photon}, as well as photon transmission through a dissipative Rydberg atomic ensemble at large input rates \cite{bienias2020photon}. On the companion side, theoretical treatment of multiphoton scattering in a Rydberg atom ensemble is rather challenging, and several methods are developed to uncover the underlying many-body dynamics. In 2013, Gorshkov, Nath, and Pohl introduced a time-ordering method \cite{gorshkov2013dissipative} to describe scattering of a short pulse in a dissipative Rydberg medium and show that the output single photon is impure. This method is since generalized to the continuous wave limit with linear EIT loss \cite{zeuthen2017correlated} and to a situation with arbitrary scattering coefficients \cite{yang2020atom}. In 2016, Gullans {\it et al}. establish an effective field theory to describe many-body dynamics in a dispersive Rydberg medium \cite{gullans2016effective}. Numerical methods such as matrix product states are also applied to the calculation of scattering dynamics \cite{bienias2020photon}.

\section{Summary and outlook}\label{sec:sec6}
This concise review briefly summarizes the basic working mechanisms and current state of the art for Rydberg atom based quantum computation and quantum simulation, covering both aspects for neutral atom and photonic QIP platforms. In particular, we highlight several recent experimental progresses within the field over the past few years, such as preparation of large-scale defect-free atomic arrays, realization of high-fidelity quantum gates, simulation of quantum spin models, and demonstration of single-photon level optical nonlinearity. These achievements pave the way towards continued successes in Rydberg atom based QIP and bring exciting prospects for developing scalable quantum computation and simulation in the coming decades.

For neutral atom based quantum computation and simulation, one important future direction is to improve fidelity for quantum state manipulation. To this end, more attention should be paid to upgrading current experimental techniques, such as examination and mitigation of error \cite{de2018analysis,levine2018high}, magic trapping of Rydberg atoms \cite{wilson2019trapped,barredo2020three}, and application of circular Rydberg states \cite{facon2016sensitive,signoles2017coherent,nguyen2018towards,aliyu2018transport,cortinas2020laser}. Continued improvements in fidelity, together with persistent efforts to increase the scale of qubit arrays \cite{de2019defect}, would enable us to achieve a Rydberg quantum computer/simulator that can outperform the best classical computer in terms of certain computing tasks.

In addition to quantum computing, neutral atom system with Rydberg interactions can be used to prepare highly entangled states such as the recently demonstrated efficient generation of $20$-qubit GHZ state through quasi-adiabatic driving \cite{Omran2019}. In the near future, it is possible to continuously increase entanglement and scale of highly nonclassical states via nonadiabatic driving \cite{Ostmann2017}, Rydberg dressing \cite{gil2014spin}, dissipation assisted evolution \cite{carr2013preparation}, quantum cellular automata paradigms \cite{Wintermantel2020}, or variational quantum approaches \cite{kaubruegger2019variational}. Meanwhile, applying these entangled states to Rydberg based precision measurement such as detection of weak electric \cite{facon2016sensitive,arias2019realization} or microwave fields \cite{sedlacek2012microwave,jing2020atomic} to enhance detection sensitivity would also be a promising direction.

For Rydberg atomic ensemble based photonic QIP, one of the urgent tasks is to overcome the limitation imposed by small blockade optical depth $d_b$ in current experiments. On the technical side, improvement can be made by increasing atomic density of the ensemble without introducing additional decoherence mechanisms \cite{gaj2014molecular,goldschmidt2016anomalous}, by employing storage enhanced nonlinearity \cite{distante2016storage,distante2017storing}, and by adapting to optical waveguides \cite{langbecker2017rydberg} or cavity \cite{parigi2012observation} setups. Meanwhile, developing novel schemes to improve scaling of the fidelity on $d_b$ holds promise for further breaking the limit caused by a finite $d_b$, e.g., utilizing dipolar exchange interaction \cite{thompson2017symmetry} or cooperative spontaneous emission in ordered atom arrays \cite{asenjo2017exponential,bekenstein2020quantum}.

Finally, we remark that hybrid QIP system involving Rydberg states is also an exciting prospect, which can integrate the advantages of different physical platforms. For example, implementing two-qubit gates in Rydberg-ion system combines state-independent trapping of ion with fast gate operation mediated by Rydberg interaction \cite{Zhang2020,mokhberi2020trapped}, and interfacing superconducting circuit with Rydberg atom provides another route towards high-fidelity gate operation together with a stable quantum information storage \cite{petrosyan2008quantum,patton2013ultrafast,hogan2012driving,morgan2020coupling,hermann2014long,hattermann2017coupling}.

\begin{acknowledgments}
This work is supported by the National Key R\&D Program of China (Grants No. 2018YFA0306504 and No. 2018YFA0306503), by the Key-Area Research and Development Program of GuangDong Province (Grant No. 2019B030330001), and by the National Natural Science Foundation of China (NSFC) (Grants No. 91636213, No. 11654001, No. 91736311, No. 91836302, and No. U1930201). L.Y. also acknowledges support from Beijing Academy of Quantum Information Sciences (BAQIS) Research Program (Grant No. Y18G24).
\end{acknowledgments}

\bibliography{ref}
\end{document}